\documentclass[12pt]{article}
\usepackage{graphics, color, soul}
\usepackage{graphicx}
\usepackage{amssymb}

\usepackage{lmodern,mathtools}

\usepackage{booktabs}
\usepackage[english]{babel}
\usepackage{amsmath,amssymb,amsbsy,amstext, amsthm, simplewick}
\usepackage{tikz}

\usepackage{graphics, color}
\usepackage{graphicx}
\usepackage{amssymb}
\usepackage{amsmath}
\usepackage[hidelinks]{hyperref}

\pdfoutput=1

\def\gsim{\, \rlap{$>$}{\lower 1.1ex\hbox{$\sim$}}\,}
\def\lsim{\, \rlap{$<$}{\lower 1.1ex\hbox{$\sim$}}\,}

\newcommand{\comment}[1]{}

\newcommand{\mbf}[1]{\mathbf #1}

\def\x{\mbf x}

\def \k {\mbf k}

\newcommand{\be}{\begin{equation}}
\newcommand{\ee}{\end{equation}}
\newcommand{\bea}{\begin{eqnarray}}
\newcommand{\eea}{\end{eqnarray}}

\begin{document}

%Title page

\begin{titlepage}

\setcounter{page}{1} \baselineskip=15.5pt \thispagestyle{empty}

\vbox{\baselineskip14pt
%\hbox{hep-th/0000000}
}
{~~~~~~~~~~~~~~~~~~~~~~~~~~~~~~~~~~~~
~~~~~~~~~~~~~~~~~~~~~~~~~~~~~~~~~~
~~~~~~~~~~~ }

\bigskip\

\vspace{2cm}
\begin{center}
{\fontsize{19}{36}\selectfont  \sc
Primordial Black Holes from non-Gaussian tails 
%{\color{red} Preliminary draft}
%Parametric Improvement of Non-Gaussianity Analysis from the Primordial Wavefunction  
}
\end{center}

\vspace{0.6cm}

\begin{center}
{\fontsize{13}{30}\selectfont   George Panagopoulos and  
Eva Silverstein}
\end{center}

%\vspace{0.2cm}

\begin{center}
\vskip 8pt

\textsl{
\emph{Stanford Institute for Theoretical Physics, Stanford University, Stanford, CA 94306}}

%\textsl{
%\emph{$^b$Institute for Advanced Study}}

%\vskip 7pt
%\textsl{ \emph{$^2$Kavli Institute for Theoretical Physics, University of California, Santa Barbara, CA 93106}}

%\vskip 7pt
%\textsl{ \emph{$^2$SLAC National Accelerator Laboratory, 2575 Sand Hill, Menlo Park, CA 94025}}

%\vskip 7pt
%\textsl{ \emph{Kavli Institute for Particle Astrophysics and Cosmology, Stanford, CA 94025}}

\end{center}

\vspace{0.5cm}
\hrule \vspace{0.1cm}
{ \noindent \textbf{Abstract}   
\vspace{0.1cm}

We develop a primordial black hole (PBH) production mechanism, deriving non-Gaussian tails from interacting quantum fields during early universe inflation.   The multi-field potential landscape may contain relatively flat directions, as a result of energetically favorable adjustments of fields coupled to the inflaton.   Such additional fields do not contribute to CMB fluctuations given a sufficient large-scale decay, related to a gap in the critical exponents computed using stochastic methods.  
Along such directions transverse to the inflaton, the field makes rare jumps to large values. 
% The dominant spatial scale of such jumps is over a Hubble scale.   
Mixing with the inflaton leads to a substantial tail in the resulting probability distribution for the primordial perturbations.  Incorporating a large number of flavors of fields ensures theoretical control of radiative corrections and a substantial abundance.  This generates significant PBH production for a reasonable window of parameters, with the mass range determined by the time period of mixing and the inflationary Hubble scale.  We analyze a particular model in detail, and then comment on a broader family of models in this class which suggests a mechanism for primordial seeds for early super-massive black holes in the universe.  Along the way, we encounter an analytically tractable example of stochastic dynamics and provide some representative calculations of its correlations and probability distributions.  

\vspace{0.3cm}

 \hrule

\vspace{0.6cm}}
\end{titlepage}

\tableofcontents

\section{Introduction}

The possibility   that some black holes in the universe might originate from primordial fluctuations \cite{Carr}\ remains of significant interest (see e.g. \cite{otherearly}\ for other early studies and \cite{recentreviews}\ for recent reviews).    
The production of primordial black holes (PBH) requires large initial perturbations $\zeta\gtrsim 1$, yielding sufficient overdensities to collapse when they re-enter the horizon after inflation.\footnote{There are more precise calculations such as \cite{zetacrit} of the conditions on $\zeta$ and related variables, with interesting subtleties discussed recently in \cite{Kopp}\cite{Caneretal}\ and references therein.}  The large-$\zeta$ tail of the probability distribution for $\zeta$ determines the abundance of black holes, with their mass determined by its physical scale of correlation.

In another line of development, the study of primordial non-Gaussianity yields concrete constraints on interacting quantum fields in the early universe \cite{Gdata}.  Although much of the effort there has focused on the size and shape of low point correlation functions of $\zeta$ \cite{shapes}, the full probability distribution plays an important role in some scenarios \cite{BondBilliards}\cite{AndreiWeb}\cite{PajerN}\cite{productive}\cite{Bruno}.  We lack a systematic understanding of non-Gaussianity since the probability distribution for the fields in the early universe  can take many forms.              

In this work, we develop a PBH formation mechanism arising from multifield dynamics in the early universe, with additional field sectors mixing with the inflaton fluctuations in a simple and theoretically natural way.  This leads to a strongly non-Gaussian tail for $\zeta$, with black holes generated at a mass scale which is determined by the inflationary Hubble scale and the time period of mixing.   The work \cite{Starobinsky}\ derived the probability distribution for field fluctuations in separate patches of the universe, whose size is given by a correlation scale determined in terms of model parameters.  This in particular allows us to derive the decay of the additional fluctuations on CMB scales.  We apply this to a range of models, including those with relatively flat potentials for additional light fields. This can come for example from the adjustment of heavy fields \cite{flattening}, or other effects such as nontrivial kinetic terms \cite{alpha}.  As an offshoot of our analysis, we encounter a model whose stochastic dynamics is analytically tractable, which we analyze in some detail in the main text and appendices.       

%This yields a probability distribution for the field $\chi$ in approximately independent patches of the form $e^{-4\pi^2V(\chi)/3 H^4}\sim e^{-\lambda_p\chi^p}$ with $\hat p<2$.    

Prescribing a simple mixing interaction with the inflaton fluctuations 
%around the reheating time 
yields strong non-Gaussian tails in the resulting distribution for the scalar perturbation $\zeta$.  As we will see, simple examples with potential drift toward the origin in the additional field space yields a non-Gaussian distribution of the form $e^{-(\zeta/\tilde\sigma)}$.     
This non-Gaussian tail in turn generates a sizable PBH density, for a reasonable window of couplings given a sufficiently large number of field flavors $N_f$ (which we introduce also for control of radiative corrections).    A similar distribution was obtained in the interesting scenarios \cite{otherlinear}, whose radiative stability would be interesting to investigate further.      We argue that certain well-motivated generalizations, which do not require a large number of fields, will produce even stronger non-Gaussian tails, and we comment on this as a candidate for super-massive black hole seeds.

%  with reasonable values of the couplings and field number, not requiring extreme values or any special features in the potential or interactions.      

%In this note, we focus on a simple example of this general scenario and make very basic estimates for PBH physics.   

We will present additional theoretical background and results on high N point correlation functions in another work \cite{uslong}, and we leave a more detailed study of the phenomenology for future work.     

\section{Probability distribution and correlation length}

Let us consider a second light field $\chi$, beyond the inflaton field $\phi$  with action 
\be\label{chiLag}
{\cal S}_\chi=\int d^4 x \sqrt{g}\left\{ \frac{1}{2}g^{\mu\nu} \partial_\mu\chi \partial_\nu \chi -V(\chi) \right\}
\ee
during inflation, where $g_{\mu\nu}$ is the inflationary metric, approximately $-dt^2+e^{2 Ht} d\x^2$.     
For simplicity, we take $V(\chi)$  to be subdominant to the inflationary potential in driving inflation.  
Although we will set up the problem with two light fields, the scenario generalizes readily to larger numbers of fields as we will see below.  

On the approximate de Sitter solution that pertains during inflation, it was shown in \cite{Starobinsky} that the field evolves toward an equilibrium configuration with a correlation length depending on $V(\chi)$.   We will draw from this work, and refer the reader to \cite{Starobinsky}\ for detailed derivations.    The field may be separated into long and short modes, with the former evolving like Brownian motion in a potential, with noise arising from the quantum fluctuations of the field.  On independent patches of size given by a correlation length $R_c$, the field is drawn from a distribution
% {\color{red}  Factor of two error here, I have traced it through in red below}
\be\label{rhoeq}
\rho_{eq}\sim {\cal N}_{eq} e^{- 8\pi^2 V(\chi)/3 H^4}
\ee
with ${\cal N}_{eq}$ the normalization constant.  
The correlation length is given roughly by
\be\label{RS}
R_c \sim H^{-1}e^{ H/\Lambda_1}.
\ee
where $\Lambda_1$ governs the rate of approach to equilibrium, with the leading non-equilibrium contribution to the full distribution $\rho(t)$ scaling like $e^{-\Lambda_1 t}$.  This rate $\Lambda_1$ is in turn determined by the function $V(\chi)$.   For the study of primordial black holes, we will be interested in the regime where this correlation length fits well within the observed universe:
\be\label{Rcupperbound}
R_c\ll e^{N_e}H^{-1} \Rightarrow \frac{\Lambda_1}{H } > \frac{1}{N_e}
\ee
where $N_e$ is the number of e-foldings in our observed patch.  In particular, we will be concerned with the falloff of the contributions of $\chi$ to the curvature perturbations on CMB scales.  

In more detail, these statements follow from the analysis \cite{Starobinsky}\ of 
a Fokker-Planck equation for the one-point pdf of sufficiently long modes of $\chi$, which takes the form
\be\label{FPone}
\frac{\partial\rho_1}{\partial t}=\frac{H^3}{8\pi^2}\frac{\partial^2\rho_1}{\partial\chi^2}+\frac{1}{3 H}\frac{\partial}{\partial\chi}\left({V'(\chi)\rho_1}\right)
\ee
The equilibrium solution (\ref{rhoeq}) satisfies $\partial \rho_1/\partial t =0$.  To study more general solutions, it is useful to work with a basis of energy eigenstates of an analogue Schrodinger problem \cite{Starobinsky}
\be\label{StarSchrod}
\left( -\frac{\partial^2}{\partial\chi^2} + [v'(\chi)^2-v''(\chi)]\right)\Phi_n(\chi)= \left(-\frac{\partial}{\partial\chi}+v'(\chi)\right) \left(\frac{\partial}{\partial\chi}+v'(\chi)\right)
\Phi_n(\chi)=\frac{8\pi^2\Lambda_n}{H^3}\Phi_n(\chi)
\ee
with 
\be\label{vsmall}
v(\chi)=4\pi^2 V(\chi)/3 H^4.
\ee
The effective potential $w(\chi)\equiv  v'(\chi)^2-v''(\chi)$ in this problem leads to a vanishing lowest eigenvalue, $\Lambda_0=0$; this corresponds to the equilibrium solution $\propto e^{-v(\chi)}$ (\ref{rhoeq}) as can be seen immediately from the middle form of (\ref{StarSchrod}).  The solutions $\Phi_{n>0}$ with nonzero eigenvalues $\Lambda_{n}>0$ of (\ref{StarSchrod}) lead to exponentially decaying contributions to $\rho$, with $\rho_n$ depending on time as $e^{-\Lambda_n t}$.
%$\Lambda_{n>0}$ are gapped, as we approach equilibrium the non-equilibrium terms are suppressed exponentially, $\sim e^{-\Lambda_n (t-t_0)}$. 
    
The authors \cite{Starobinsky}\ also generalized their analysis to compute the joint distribution  two point function $\langle\chi(\x, t)\chi(\x', t')\rangle$, using the Fokker-Planck equation for the joint probability distribution of the field at two separate points.   This leads to a two point function of the field in this equilibrium distribution which we briefly review in the appendix.  Roughly speaking, it behaves at equal times $t$ and {\it large} proper spatial separation $R=a(t)|\Delta \x|$ as         
\be\label{GRgen}
G(R)\equiv \langle \chi(\x_0, t)\chi(\x_0+\Delta \x, t)\rangle \sim \frac{H^2}{(R H)^{2 \Lambda_1/H}}
\ee
plus contributions of higher modes that are suppressed by larger exponents $\Lambda_n$.   In some cases, such as the $\lambda\chi^4$ theory studied in detail in \cite{Starobinsky}, these additional terms can be neglected to good approximation.   We work out an example analytically in the appendix, including these additional contributions, finding that the estimate (\ref{GRgen}) remains a reasonable approximation parametrically.         
At zero separation, on the other hand, we have a two point function 
\be\label{Gzero}
G(0) =  \int d\chi ~\rho_{eq}(\chi) ~\chi^2
\ee
%The proportionality constant depends on model parameters but is independent of $R$.  
Defining a correlation length $R_c$ by the scale at which $G(R)/G(0)$ is suppressed by a  constant factor $e^2$ yields (\ref{RS}).
%\footnote{If the factor is $\eta_c$ (with $\eta_c=2$ taken in \cite{Starobinsky}), then $\gamma_c=\log(\eta_c)/2$.}  
%We will illustrate this explicitly in a family of models shortly.  

For us, this will be important in estimating
%the ratio of $G(R_c)$ to 
the two point correlation function at CMB scales:
\be\label{CMBBHRatio}
{G(R_{CMB})}  \sim H^2\left(\frac{1}{R_{CMB}H}\right)^{2\Lambda_1/H}
\ee
{
%\cyan I think this is not exactly the power-spectrum. Maybe we could say 
For $k/a \ll 1/R_c$ the power spectrum is
\be
P(k) \sim \frac{H^2 (-k/aH)^{2\Lambda_1/H}}{k^3}
\ee
Hence, at the end of inflation when $R_{CMB}=a/k_{CMB} \gg R_c$, it is suppressed by $(R_{CMB} H)^{-2\Lambda_1/H}$.}
For values of $\Lambda_1/H$ of interest to us for PBH formation, the ratio $R_{CMB}/R_c$ will be very large, which goes in the direction of suppressing its effect on CMB scales.  Still, once we incorporate mixing interactions with the inflaton fluctuations, we must check that the contributions of our $\chi$ sector to the CMB correlations will be negligible.  The $\chi$ sector contributes blue perturbations, with spectral index $\sim \Lambda_1/H$.    

All of this can be studied in great generality, and we can consider a wide variety of models.  To formulate one family of models, we will take into account a particular property $V(\chi)$ can have in a multifield landscape, articulated in \cite{flattening}.  Generically in field theory, and its ultraviolet completions, we may expect interacting fields of various masses.  Couplings between our light field $\chi$ and heavier ones leads to energetically favorable adjustments of the heavy fields as the system builds up potential energy in the $\chi$ direction.  This is a simple, classical effect which typically leads to potentials $V(\chi)$ which are flatter than quadratic.   
%To be specific, we will focus on a representative example from a simple family of models which are sufficient to illustrate the effect.  
Let us consider potentials (\ref{vsmall}) which take the form
\be\label{Vpowers}
v(\chi)  \to \lambda_p\left|\frac{\chi}{H}\right|^p
\ee
for $\chi$ greater than some scale $M$. For the physical reason just reviewed, we will consider more generic $p$ than integers.  Powers $p<2$ appeared earlier in axion monodromy models, a feature which can be understood via this flattening effect of heavy fields \cite{flattening}, but it is a more general feature of multifield potential landscapes which can be understood in simple field theory models.    For $p < 1$, the Schrodinger problem (\ref{StarSchrod}) exhibits no gap to the first excited state $\Lambda_1$; instead there is a continuum above zero, starting from scattering states at $\chi\to\infty$.  We defer this case to future work; see \cite{uslong}\ for more comments.  Here we restrict ourselves to $p\ge 1$.   The theory with $p=2$ is free in the $\chi$ sector.  We will see that the nontrivial example $p=1$ is calculable in the stochastic framework, something that might be of more general interest.  We will analyze $p=1$ and $p=2$ in detail, but as we will explain below the main results readily apply more generally for potentials with an inward drift in the $\chi$ direction.  After deriving the results for that case, we will comment on the generalization to potentials with a richer shape with both signs of drift.  

In this family of models (\ref{Vpowers}), $G(0)$ in (\ref{Gzero}) becomes
\be\label{Gzerop}
G_p(0)= H^2\lambda_p^{-2/p} \frac{\Gamma(\frac{3+p}{p})}{3\Gamma(1+\frac{1}{p})}
\ee            
For example for $p=1$, this reduces to $G_1(0)=2H^2\lambda_1^{-2}$.  Computing $\Lambda_1$ using the Schrodinger problem above yields 
\be\label{Lamoneone}
\frac{\Lambda_1}{H} =  \frac{\lambda_1^2}{8\pi^2}, ~~~~~ p=1.
\ee
There is a continuum of functions $\Phi_{\pi_\chi}=\exp(i \pi_\chi \chi) $ solving (\ref{StarSchrod}) with eigenvalues above this gap:
\be\label{poneLambda}
\frac{\Lambda}{H}=\frac{H^2}{8\pi^2}\left(\pi_\chi^2+\frac{\lambda_1^2}{H^2}\right), ~~~~ p=1. 
\ee
This model is analytically solvable.
We work out its two point function (along with its joint probability distribution on two patches) for this model in the appendix, finding that it essentially behaves as in the simple estimate (\ref{GRgen}).    
For the trivial $p=2$ theory, the corresponding spectrum is
\be\label{Lambdanp2}
\Lambda_n = \frac{\lambda_2}{2\pi ^2}n, ~~~~~~~~~~~~~~ p=2
\ee

%Let us also spell out the estimate of the correlation length $R_c$ in this example, following \cite{Starobinsky}:
%\be\label{GRGzero}
%2 \equiv \frac{G(0)}{G(R_c)}\sim (H R_c)^{c_\Lambda \lambda_1^2/H^6}2 \left(\frac{\lambda_1}{H^3}\right)^{-2}
%\ee      

So far, we have focused on the $\chi$ sector evolving independently in the inflationary near-de Sitter solution. 
The state of $\chi$ and $\delta\phi$ before the exit from inflation takes the product form $\Psi_G(\delta\phi)\Psi_\perp(\chi)$, where $\Psi_G$ is the Gaussian wavefunction for the leading inflaton fluctuations, and $\Psi_\perp(\chi)$ is the wavefunction for $\chi$ in the equilibrium distribution just described.  In general, this and the corresponding probability $\rho(\chi(x))=|\Psi(\chi(x))|^2$ are complicated functionals of the field.   The probability reduces to the equilibrium distribution (\ref{rhoeq}) described above on roughly independent patches of size $R_c$. The one-point pdf $\rho_1$ (\ref{rhoeq}), i.e. the histogram of $\chi$ excursions in a volume, is in general strongly non-Gaussian, as are the higher-point joint probability distributions.   

{
However, for our present purposes these distributions are not directly relevant and the problem simplifies for potentials $V(\chi)$ which have an inward drift (including all quantum corrections).  The nonlinearities in $\rho_N$ arise via sufficiently long time-evolution of modes outside the horizon, up to a timescale of $N_e$ for the longest modes which exit the horizon first. The shortest modes which exit the horizon near the end of inflation remain Gaussian.  For primordial black hole production, we will be interested in large values of the curvature perturbation $\nabla^2\zeta$, and this will correspond to rare large excursions of $\chi$.  A long mode with extensive time evolution would roll down a potential with inward drift, reducing the value of $\chi$.  Long modes also have smaller gradients, contributing less to the curvature, and they have a smaller variance in field space.  Shorter modes remain very close to their initial excursion, aside from doubly-rare events in which a shorter mode kicks it to the other side of the potential.  As a result, the modes relevant for black holes are determined by a nearly Gaussian distribution for $\chi$.   This holds in particular for the family of models  (\ref{Vpowers}).

We will comment further below on potentials containing regions with outward drift.  This includes symmetry-breaking potentials as well as potentials with oscillatory features.  

}

Over a range of possible time periods during inflation or afterwards,  new interactions can develop among the fields, including couplings between $\chi$ and $\phi$.  This could happen during the exit from inflation and reheating, or at other times.  We will analyze a specific, calculable class of models of this kind shortly.  Before specializing to that, let us consider the generic case to gain some additional perspective.  
Consider a system that is suddenly coupled locally to another system at a time $t_{mix}$,
without tuning any couplings to be small.  Suppose we start in the ground state, a product of Bunch-Davies Gaussian states $\psi_G(\delta\phi)\otimes\psi_G(\chi)$ as we will do in our model.  In the sudden approximation of quantum mechanics, this is the state just after $t_{mix}$, and then it evolves by the full (mixed) Hamiltonian:
\be\label{expandstate}
\Psi(t)=\sum_n c_n e^{-i E_n (t-t_{mix})} |\psi_n\rangle, ~~~~ c_n=\langle \psi_G(\delta\phi)\otimes\psi_G(\chi)| \psi_n\rangle
\ee
where $|\psi_n\rangle $ are the energy eigenstates of the in general strongly mixed Hamiltonian, with eigenvalues $E_n$.   Now
\be\label{tracechi}
|\Psi(\delta\phi_0, \chi_0, t_0)\rangle \langle \Psi(\delta\phi_0, \chi_0, t_0)| = \sum_{n, n'} c_n c^*_{n'} e^{i (E_{n'}-E_n)(\Delta t)}|\psi_n\rangle \langle \psi_{n'} |
\ee
and for $\Delta t\gg 1/\Delta E$ this will collapse onto the diagonal, giving a probability distribution for $\delta\phi$ of the form 
\be\label{rhodiag}
\rho(\delta\phi_0)\simeq  \int_\chi \sum_n |c_n|^2 |\psi_n(\delta\phi_0, \chi)|^2
\ee
These energy eigenstates $\psi_n$ of the new system will be highly mixed states, far from the initial product of Gaussians, with strong entanglement.   This may also lead to a non-Gaussian tail in the distribution (\ref{rhodiag}).  Of course,  it is difficult to calculate this in the general case, and in this paper we will analyze a more specific scenario. 

Let us consider a mixing interaction of the form
\be\label{mixgen}
{\cal S}_{mix}=\int dt \int d\x a(t)^3 F_{mix}(\chi) \dot\phi^2 
\ee
pertaining over a timescale 
\be\label{Deltatineq}
\Delta t \ll N_e H^{-1}
\ee
starting at some time $t_{mix}$ which may be during or after inflation.  We imagine this arises from rolling scalar fields, some of which behave like time-dependent effective couplings; it would be interesting to model this in more detail.

Similarly to our discussion of the potential, one could consider a variety of possibilities for $F(\chi)$, but we will focus on the simple case
\be\label{mixUV}
F_{mix}(\chi)= \left(\frac{\chi}{M_*}\right)^m 
\ee
We can consider, for example, $m=1$ or $2$, depending on the symmetry structure of the theory, or potentially other values analogously to the discussion of $p$ above.  We will focus on the case $m=2$, respecting a reflection symmetry in $\chi$ field space.   Here $M_*$ is an energy scale, one of the parameters which will enter into the phenomenology of the model.  In the appendix, we derive a window of parameters for which the leading radiative correction to the effective action of the $m=2$ model is computable (as in (\ref{actionfull})) and subdominant to the contributions we will focus on in the main text.  This involves a large flavor expansion, so for the $m=2$ model we will introduce a number $N_f\gg 1$ of flavors of $\chi$ fields respecting an $O(N_f)$ global symmetry, denoting them collectively by $\vec\chi$.      
Finally, after working out the dynamics, radiative stability, and predictions of the $m=2$ model and its multifield generalization, we will discuss an interesting generalization with $F(\chi)\propto \exp(2\chi/M_*)$.\footnote{This is motivated in part by the exponential relation between the dynamical couplings and length scales in string theory and the corresponding canonically normalized scalar fields:  the string coupling takes the form $g_s\sim e^{\sigma_c/M_p}$ in terms of the canonically normalized field $\sigma_c$.}  

%However, we note that the interaction with the rolling inflaton field (with $\dot\phi^2\sim \epsilon M_{Pl}^2 H^2$ in terms of the inflationary slow roll parameter $\epsilon$) may flatten the function $F(\chi)$ analogously to the flattening of the potential $V(\chi)$ discussed above.      

We will be able to understand the strength and effect of this mixing interaction in a simple way after a few more steps.  First, we note that during inflation, $\dot\phi=\dot{\bar\phi}(t)+\delta\dot\phi(\x, t)$, with the leading, homogenous piece related to the amplitude of the power spectrum as $\dot{\bar\phi}/H^2\sim 10^5$.  During the exit and reheating phase, as well as for sufficiently brief periods during inflation, 
$\dot{\bar\phi}$ may exceed this.  
The leading contribution to the mixing is given by replacing one of the $\dot\phi$ factors in (\ref{mixgen}) with $\dot{\bar\phi}$, giving
\be\label{mixnext}
{\cal S}_{mix}\sim \int dt \int d\x \dot{\bar\phi} [a(t)^3\delta\dot\phi] F(\vec\chi)
%\left(\frac{\chi}{M_*}\right)^m  
\ee
The quantity in square brackets is the conjugate momentum to the inflaton fluctuation $\delta\phi$,  $\Pi_{\delta \phi}= a(t)^3 \delta\dot\phi$, so in canonical variables we have a mixing interaction ${\cal H}_{mix}=\int d\x~ \dot{\bar\phi}~\Pi_{\delta\phi}F(\chi)$.  The evolution of the state of system under the mixing interaction in the time $\Delta t$ can be usefully written (as in ordinary quantum mechanics) as Hamiltonian evolution of the state
\be\label{Hform}
|\Psi ( t_{exit}+\Delta t)\rangle \simeq e^{-i \Delta t {\cal H}} |\Psi\rangle \simeq e^{- i \Delta t \dot{\bar\phi} \int d\x \Pi_{\delta\phi}F_{mix}(\chi)} |\Psi\rangle 
\ee
We are neglecting the free evolution of the state, which is indeed highly subdominant for superhorizon modes, since
\be\label{freeH}
H_{free}\sim \int d\x \frac{{ \Pi_{\delta\phi}^2 +\Pi_\chi^2}}{ a(t)^3}
\ee
is exponentially suppressed at late times compared to the mixing interaction. 

In this form, we can recognize the evolution operator as the generator of translations in field space: 
\be\label{phitrans}
\Psi (\vec\chi, \delta\phi, t_{exit}+\Delta t) = \Psi(\vec\chi, \delta\phi+\dot{\bar\phi}\Delta t F_{mix}(\vec\chi), t_{exit})
\ee
Let us now restrict attention to the example (\ref{mixUV}) with $m=2$ and define
\be\label{kappadef}
\kappa = \frac{\Delta t \dot{\bar\phi}}{M_*^2}\equiv \Delta t\lambda_{mix}
\ee
In the appendix we will discuss further the theoretically consistent range of values of $\lambda_{mix}$, finding viable windows of parameters consistent with radiative stability of the model.

%{The size of this mixing interaction will enter into our black hole abundance.  It comes naturally enhanced by a factor of $\frac{\dot{\bar\phi}}{H^2}\sim 10^5$.  However, its size may be limited by theoretical control.   If we assume that the scale $M_*$ governs corrections that arise as powers of both $\chi/M_*$ and $\frac{\dot\phi}{M_*^2}$, the simplest regime to work in is one in which these are kept $\ll 1$ so that we can safely ignore both classical and quantum corrections that arise as higher powers of these quantities. This would in particular require
%\be\label{perturbativityI}
%\frac{\dot\phi}{M_*^2}\ll 1 \Rightarrow M_* \gg 10^{5/2}H
%\ee
%This in turn imposes an upper bound on kappa, which becomes  using (\ref{kappadef})
%\be\label{perturbativityII}
%\frac{\dot\phi}{M_*^2}\ll 1\Rightarrow \kappa = (H\Delta t) \frac{\dot{\bar\phi}}{H^2}\left(\frac{H}{M_*}\right)^m\sim (H\Delta t)10^5\frac{H^2}{M_*^2}\ll N_e 10^{5-5m/2}
%\ee
%where the last expression takes into account that $H\Delta t < N_e$.  For $m=1$ this would give $\kappa\ll N_e 10^{5/2}$, and for $m=2$ it gives $\kappa \ll N_e$.   
%
%However, in analyzing our examples of PBH production below, we will also explore values of $\kappa$ beyond this range.  In the appendix, we derive a window of couplings, including an additional flavor number $N_f$ of our $\chi$ sector, for which we can control the radiative corrections to the model while obtaining a substantial PBH abundance.    

The state of $\vec\chi$ and $\delta\phi$ following the mixing takes the form
\be\label{Psiafter}
\Psi(\vec\chi, \delta\phi)=\Psi_G(\delta\phi+\kappa \vec\chi^2) \Psi_\perp(\vec\chi) 
\ee
Finally, the probability distribution for $\delta\phi(\x)$ is given by squaring the wavefunction and tracing over $\vec\chi$, $\int D\vec\chi |\Psi(\vec\chi,\delta\phi)|^2$.  At the level of the pdf for individual patches of size $\sim H^{-1}$, this translates to
\be\label{rhodelphi}
\rho(\delta\phi)=\frac{1}{\sqrt{2\pi}\sigma}\int d\vec\chi \rho_{G}(\vec\chi) e^{-(\delta\phi-\kappa \vec\chi^2)^2/2\sigma^2}
%= \frac{1}{\sqrt{2\pi}\sigma}\int d\vec\chi e^{-\vec\chi^2/2\sigma^2} e^{-(\delta\phi-\kappa \vec\chi^2)^2/2\sigma^2}
\ee
with $\sigma^2\sim  H^2/4\pi^2$.   
%over $N_e$ e-foldings of inflation, 
%where $\rho_{eq}$ was given above in (\ref{rhoeq}).  
%Here 
%\be\label{Neq}
%{\cal N}_{eq}= \left(2\int_0^\infty d\chi e^{-{\color{red}2}\lambda_p  (\frac{\chi}{H})^p} \right)^{-1}=\frac{({\color{red} 2}\lambda_p)^{1/p}}{2 H \Gamma(1+\frac{1}{p})}
%\ee
In the next section, we will integrate over the appropriate region of the tail of the distribution (\ref{rhodelphi}) to estimate the PBH abundance.  For now, let us simply note that the tail behaves like $\exp(-\delta\phi/\kappa)$, as can be seen by a simple saddle point estimate.  This is a non-Gaussian tail in $\delta\phi$.  

%make a simple estimate of the size and shape of the tail of the distribution.  
%
%The saddle point equation for the $\chi$ integral in (\ref{rhodelphi}) is
%\be\label{chisaddle}
%\kappa \chi_s^m = \delta\phi -\frac{{\color{red} 2}\lambda_p p \chi_s^{p-m}\sigma^2}{m\kappa H^p}
%\ee
%For $p<2$, and noting that $m\ge 1$, the second term on the right hand side becomes subdominant for large field values.  The field values in the regime of interest below will be sufficiently large to neglect the second term, as a result of the large value of $\delta\phi\sim \zeta \dot{\bar\phi}/H$ required to form black holes.   
%
%We then obtain a tail of the form
%\be\label{tailform}
%\rho(\delta\phi) \sim e^{-{\color{red}2}\lambda_p (\frac{\delta\phi}{\kappa})^{p/m}} 
%\ee
Converting to $\zeta\sim\delta\phi H/\dot{\bar\phi}$, this takes the form
\be\label{zetatailform}
\rho(\zeta)\sim e^{-C \zeta},  ~~~~~ C=\frac{M_*^2}{(H\Delta t)}
\ee    
%For $\hat p<1$, the tail of this distribution exceeds the Gaussian.  
For more general $m$, this would take the form $\rho\sim \exp(- C\zeta^{2/m})$.  
We note here the similarity between this distribution and the one considered in the recent work \cite{SMBHmu}, which argued that a viable mechanism for supermassive black hole seeds would require $2/m \le 1/2$.  In this class of models that would require $m=4$, which does not appear natural.  However, again we note that more general behaviors might arise from richer forms of the potential $V(\chi)$.  Moreover, we will also consider an exponential form for $F(\chi)$, which will introduce a much heavier tail consistent with the condition in \cite{SMBHmu}.

%{\cyan MM: I don't fully understand this last comment.}   

%with constant $C_\zeta$.                          
%**Tail estimate (below will integrate over $\delta\phi$, giving error function).  

%\be\label{kapsize}
%\kappa \sim  \Delta t \frac{\dot{\bar \phi}}{M_*} \sim (H\Delta t)\frac{H}{M_*} \frac{\dot{\bar \phi}}{H^2}
%\ee

\section{Primordial Black Holes}

We are now in a position to analyze the PBH production in our theory.  We will focus on the simplest estimates of mass scale and abundance, although it would be interesting to pursue more detailed studies in the future.  In this section, we lay out the formulas for PBH abundance and constraints from the CMB, for simplicity starting with a single $\chi$ sector.  We will then generalize to multiple $\chi$ fields, and exhibit substantial PBH production.  

\subsection{Mass scale of the black holes}

In our scenario, we mix $\chi$ into $\delta\phi\sim \zeta\dot\phi/H$  at some time $t_{mix}$ around the exit from inflation, as discussed above.  At that point, the one-patch pdf of the $\zeta$ probability distribution is given by (\ref{rhodelphi}) which can be traded for the curvature perturbation via $\delta\phi \sim \zeta \dot\phi/H$. Afterwards, every comoving mode of $\zeta$ remains conserved until the horizon re-entry. 

Let us consider black holes formed from $\chi$ excursions of some size $R_* a/a(t_{mix})$.   By this we mean that the field smoothed on scale $R_*$ has an excursion corresponding to an order one curvature fluctuation.  Although we argued above that at least for inward-drifting potentials, this size will be $H^{-1}$, we will first write the general formula.
%In particular, sufficiently far out in $\zeta$, its tail is dominated by the non-Gaussian contribution injected by the $\chi$ sector, which is decorrelated outside the expanding physical scale $R_c a/a(t_{mix})$.  This statement is based on the two point correlation function derived in \cite{Starobinsky}, which we will assume provides a good guide to  the scale of correlation of the full nonlinear theory.   
We expect black holes of Schwarzschild radius $r_s = 2 G_N M$ to form when this scale crosses the horizon at the time $t_h$, i.e.
\be 
%\label{labelrsRS}
\label{MS}
r_s \sim H^{-1}(t_h) \sim R_* \frac{a(t_h)}{a(t_{mix})}
%r_s \sim R_{Sh}
\ee  
%with mass
%\be\label{MS}
%M \sim R_{Sh} M_{Pl}^2.  
%\ee
%This of course assumes that $t_{mix}<t_h$.  
This allows a wide range of mass scales
\be\label{Mass}
M\sim R_* M_P^2 \frac{a(t_h)}{a(t_{mix})}\sim \frac{ M_P^2}{H} \frac{a(t_h)}{a(t_{mix})}
\ee 
obtained by varying the mixing time.  In the last relation here, we assumed an inward drifting potential with $R_*\sim H^{-1}$; this is conservative as other cases will have a larger non-Gaussian contribution from the $\chi$ sector.

In the special case that (i) there is an instantaneous transition from inflation to radiation epoch, where $a \propto t^{1/2}$, and (ii) mixing happens soon after reheating such that 
\be
H(t_{mix})=\frac{1}{2t_{mix}}\sim H,
\ee
where $H$ is the expansion rate during inflation, and (iii) horizon crossing occurs during radiation domination, we get
%the following relation between $R_S$ and $R_{Sh}$.  During this period $a(t)=a_{exit}(t/t_{exit})^{1/2}$, so $H^{-1}(t)=H^{-1}(t/t_{exit})=H^{-1}(a(t)/a_{exit})^2$.  Horizon crossing is at $H^{-1}(t_h)=R_{Sh}=R_S(a_h/a_{exit})$.  Putting this together with the previous relations gives $a_h/a_{exit}=HR_S$, and hence $R_{Sh}=H R_S^2 $.  So (\ref{MS}) becomes
\be\label{MSrad}
M \sim {R_*^2}H M_{Pl}^2 \to \frac{M_P^2}{H}, ~~~~~ R_*\to H^{-1} ~~~({\text{for}}~~ t_{mix}\sim t_{reheating})
%M \sim R_{Sh} M_P^2 \sim R_S^2 H  M_P^2 \sim e^{2 H/\Lambda_1}\left(\frac{M_P}{H}\right) M_P
\ee
In this special case, one key mass scale is $M\sim 10^{15}g$, the threshold between evaporating and non-evaporating black holes.   This corresponds to an inflationary Hubble scale of order $10^{-20}M_P$, or equivalently an energy scale of $V_{inflation}^{1/4}\sim 10^{-10}M_P$.   As noted above, earlier mixing leads to a wide range of heavier PBH masses (\ref{Mass}) for a fixed value of the inflationary Hubble scale.  

\subsection{Abundance of black holes}

The initial abundance $\beta$ of primordial black holes is given by the tail of the distribution. We will express this in terms of a critical value of $\zeta\sim \frac{H}{\dot{\bar\phi}} \delta\phi>\zeta_c$ as in \cite{zetacrit}.  We note that interesting recent work has analyzed non-Gaussian effects arising from a conversion to a different variable used to assess black hole formation \cite{Caneretal}, in the case of a purely Gaussian distribution of $\zeta$, finding that these suppress the abundance for a given fluctuation amplitude.  On the other hand, the variable $\zeta$ (related to the proper size of the excursion) gives a clearer description of the range of perturbations
and can be used to assess black hole formation \cite{Kopp}.  We will not delve into these interesting issues here, since our interest is in the primordial non-Gaussian tails which in themselves enhance the effect.  Even if some adjustment of $\zeta_c$ is needed, our mechanism will apply.  In the model developed explicitly in the main text, a viable (or larger) abundance arises in any case for an appropriate range of the number $N_f$ of flavors of $\chi$ fields.

\be\label{betadef}
\beta 
%=\frac{\rho_{PBH}}{\rho_{Tot}} 
\simeq \int_{\zeta_c\dot{\bar \phi}/H}^\infty d\delta\phi  \rho(\delta\phi)
\ee
In our scenario, this will essentially be given by integrating the distribution $\rho(\zeta)$ given above in  (\ref{zetatailform}) from $\zeta=\zeta_c$ to $\infty$, but we will analyze it in this section starting from (\ref{rhodelphi}).    

This is immediately constrained by the basic requirement that it not exceed the dark matter abundance.  As reviewed in \cite{PBHconstraints}, this condition takes the form
\be\label{basicbetaconstraint}
\Omega_{PBH}<\Omega_{CDM}\Rightarrow \beta(M) < 10^{-28} \sqrt{\frac{M}{M_{Pl}}}, ~~~~ M>10^{15}gr
\ee
up to various order one ratios that are detailed in the references (see e.g. eqn (2.6) of  \cite{PBHconstraints}). 
Here the inequality $M>10^{15}gr$ ensures that the black holes have not yet evaporated.  
This is a general requirement; we will consider the window studied in \cite{SMBHmu}\ of $10^5M_{\odot} < M < 10^{11}M_{\odot}$ as a special case below.  
For smaller masses, interesting bounds also exist, e.g. from fig 3 of \cite{PBHconstraints}\ we see that big bang nucleosynthesis restricts $\beta$ (again up to order one coefficients) as roughly
\be\label{lowerbetaconstraint}
\beta < 10^{-20}, ~~~~~~ M\sim 10^9 -10^{15}gr
\ee
There are also stronger constraints in various other mass ranges.\footnote{For a recent treatment of constraints involving extended mass functions, see e.g. \cite{extended}.}  In this paper we wish to determine whether the non-Gaussian tails generated above can produce detectable levels of PBHs, for natural ranges of parameters.  

%For that basic purpose we will mostly focus on the simple estimate (\ref{betadef}), although we will also note an intriguing connection of our models to some remarks in \cite{SMBHmu}.     

In our analysis below, will take $\zeta_c=.1$ as estimated in \cite{zetacrit}.
%this is conservative for us in determining how detectable our PBH production is.  
The abundance (\ref{betadef}) is then explicitly
\be\label{betachi}
\beta(M)=\int d\vec\chi \rho_{G}(\vec\chi) \int_{\zeta_c\dot{\bar \phi}/H}^\infty {d\delta\phi}
\frac{1}{\sqrt{2\pi}\sigma} e^{-(\delta\phi-\kappa \vec\chi^2)^2/2\sigma^2}
=\int d\vec\chi \rho_{G}(\vec\chi) \frac{1}{2}{\text{erfc}} \left(\frac{\zeta_c\frac{\dot{\bar \phi}}{H}-\kappa \vec\chi^2}{\sqrt{2}\sigma}\right)
\ee
The complementary error function is supported (taking the constant value 2) for negative argument, which corresponds to the tail 
\be\label{chitBH}
\kappa \vec\chi^2 \gtrsim \zeta_c\frac{\dot{\bar \phi}}{H} 
\ee 
or equivalently, using (\ref{kappadef}) $|\vec\chi|\ge \chi_t$, with
\be\label{chitailsimpler}
\chi_t = \frac{M_*}{(10 H\Delta t)^{1/2}}
\ee
As discussed in the appendix, we will work in the regime $\chi_t < M_*$.

\subsection{Effect on CMB scales}

Before calculating the black hole abundance in the next section, we must also check the effect of the $\chi$ sector on the scalar perturbation on CMB scales.  We have
\be\label{twopoint}
\langle\delta\phi(t, \x_1)\delta\phi(t,\x_2)\rangle= \langle\delta\phi(t, \x_1)\delta\phi(t,\x_2)\rangle_G + \kappa^2\int D\vec\chi |\Psi_\perp(\vec\chi)|^2|\vec\chi|^m(t, \x_1)|\vec\chi|^m(t, \x_2) 
\ee
where $\kappa$ was given above in (\ref{kappadef}).  For $m=1$, the second term here, coming from the $\vec\chi$ sector, is given by  (\ref{CMBBHRatio}).  Since this is blue, its contribution to the tilt must not dominate on CMB scales.  For more general $m$, it is determined similarly using the 2-patch pdf derived in \cite{Starobinsky}, as discussed in the appendix and in \cite{GS}.    
This will be an important but easily satisfied constraint for our models.

%even though our mixing parameter $\kappa$ will be somewhat large in order to generate significant PBH production.  This is reasonably natural from the effective field theory point of view, as it contains a factor of $\dot{\bar\phi}/H^2\sim 10^5$.   

%{\cyan since $\kappa$ is a model-dependent parameter, I would say ``This will be an important constraint for our models, since our mixing parameter $\kappa$ needs to be large for the model to have significant effect on PBH production''.}

Using the formalism \cite{Starobinsky}\ reviewed above, let us lay out
the condition (\ref{twopoint}) that the $\chi$ contribution to the two point function on CMB scales not dominate over the standard one $\langle \delta\phi^2\rangle \sim H^2$ coming from the inflaton sector.  Considering $m=2$, the shift in the two point function on these scales becomes  
\bea\label{GCMBexample}
\frac{\kappa^2}{H^4} G(R_{CMB}) &\sim& \kappa^2\frac{N_f}{(R_{CMB}H)^{2\Lambda_2/H}}\sim {\kappa^2}e^{-2 N_e\Lambda_2/H} \nonumber\\
&\sim& \kappa^2 N_f e^{-N_e*\lambda_1^2/4\pi^2} ~~~~~ p=1, m=2\nonumber\\
\eea
This formula is explained in the appendix, where its precise coefficient is also derived.  We note that for a wide range of theories, including $p>1$, one would have $\Lambda_2$ gapped above $\Lambda_1$, and hence greater suppression on CMB scales for $m=2$.  In the $p=1$ case, as described in the appendix there is a continuum above $\Lambda_1$, so in the last expression here we used the fact that $\Lambda_2\simeq \Lambda_1$ in that special case.  

We will impose that this contribution on CMB scales be suppressed by $1/\sqrt{N_{pix}}\sim 10^{-3}$, where $N_{pix}$ is the number of CMB pixels.\footnote{In fact, this criterion can be relaxed to $10^{-2}$, corresponding to the Planck constraints on the tilt, running, and local non-Gaussianity parameter $f_{NL}$. 
%We will apply this slightly weaker condition in the supermassive black hole range below.
}
Our condition is then that  
\be\label{CMBcondfinal}
\kappa^2 G(R_{CMB})< 10^{-3} H^2,
\ee
with $G(R_{CMB})$ computed explicitly in the appendix.  

\section{Multifield mixing scenario and copious PBH production}

Now we will come to our estimates for PBH abundance.  
%{\color{red} Need to update factor of 2 in this section if we keep it.}
For simplicity, in order to obtain analytic results we will start with a very symmetric theory of $N_f$ fields, with potential $v(\vec\chi)=\lambda_2\vec\chi^2$ (i.e. $p=2$) and an $m=2$ mixing interaction.    The normalized one-point distribution for (\ref{rhodelphi}) becomes 
%(setting $H=1=\sigma$ for simplicity)
\be\label{rhodelphiNf}
\rho(\delta\phi)
%=\frac{1}{\sqrt{2\pi}}\int d^{N_f}\vec\chi \rho_{G}(\vec\chi) e^{-(\delta\phi-\kappa F(\vec\chi))^2/2}
= \frac{1}{\sqrt{2\pi\sigma^2}}\frac{1}{(\sqrt{2\pi\sigma^2\gamma})^{N_f}}
\int d^{N_f}\vec\chi  e^{- \frac{\vec\chi^2}{2\sigma^2\gamma}} e^{-(\delta\phi-\kappa \vec\chi^2)^2/2\sigma^2}
\ee
where
\be\label{sigmavalue}
\sigma\simeq \frac{H}{2\pi}
\ee
and $\gamma$ is a parameter that depends on the $\chi$ mass squared $m_\chi^2=\frac{3}{2\pi^2}H^2\lambda_2 $ (where we used (\ref{vsmall}).   Specifically, this depends on $\lambda_2$ as follows.  The massive mode solutions decay during the time period $\Delta t$ by an amount
\be\label{decay}
\frac{\chi_{t_m+\Delta t}}{\chi_{t_m}}=e^{-(H\Delta t)\frac{3}{2}\left(1-\sqrt{1-\frac{4 m_\chi^2}{9 H^2}}\right)}= e^{-(H\Delta t)\frac{3}{2}\left(1-\sqrt{1-\frac{2 \lambda_2}{3 \pi^2}}\right)}\simeq \frac{\chi_{t_m+\Delta t}}{\chi_{t_m}}=e^{-(H\Delta t)\frac{\lambda_2}{2\pi^2}}\simeq \sqrt{\gamma}
\ee
%We include this for completeness, but we will find that it is reasonable to set $\gamma\simeq 1$ in our analysis because our constraints will be satisfied with
where we used that the first term in the expansion of the square root is good approximation in the regime we will consider below.  In the last step, we can identify this ratio with the factor $\sqrt{\gamma}$, since $\gamma$ is the suppression in the variance of $\chi$ compared to a massless field with variance $\sigma^2$.

Let us change variables to $r^2=\vec\chi^2/\sigma^2$.  Then, doing the angular integral (using that the area of a unit sphere embedded in $d$ dimensions is $\int\Omega_d = 2\pi^{d/2}/\Gamma(d/2)$:
\be\label{rhodelphiNfradial}
\rho(\delta\phi)=\frac{2\pi^{N_f/2}}{\Gamma(N_f/2)}{\sqrt{1/2\pi}^{N_f+1}}\frac{1}{\sigma\gamma^{N_f/2}}\int dr r^{N_f-1}  e^{- r^2/2\gamma} e^{-(\delta\phi-\kappa r^2\sigma^2)^2/2\sigma^2}
\ee
We see here that for large $N_f$, the measure factor $r^{N_f}$ suppresses the origin in favor of the tails.  

Computing the abundance from this as above (cf (\ref{betachi})) by integrating over $\delta\phi$ from $\phi_c\equiv \zeta_c \dot{\bar\phi}/H$ to $\infty$ yields, after changing variables to $y=r^2$:
\be\label{betaNf}
\beta=\frac{(2\gamma)^{-N_f/2}}{\Gamma(N_f/2)}\int_{\phi_c/\kappa\sigma^2}^\infty dy y^{N_f/2-1} e^{-y/2\gamma} = \frac{\Gamma(\frac{N_f}{2},\frac{\phi_c}{2\kappa\sigma^2\gamma})}{\Gamma(\frac{N_f}{2})}
\ee
There is a saddle point for this integral at $y_s\simeq \gamma N_f$.   This saddle point is only on the contour of integration if 
\be\label{ysaddlecond}
y_s \simeq \gamma N_f > \frac{\phi_c}{\kappa \sigma^2} 
\ee
Let us consider this regime first, taking $N_f$ sufficiently large for it to hold.  (We will check below that this is consistent with the CMB constraint.)  In this case, the abundance $\beta$ is extremely large, since the integral is well approximated by sending its lower limit of integration to zero, including all $\phi$ (since the measure suppresses small $\phi$ anyway).  In that limit, $\beta\to 1$.  So the result so far is a gigantic PBH abundance
\be\label{betahuge}
\beta \lesssim 1, ~~~~~~~~~~ {\text{assuming}}~~ N_f > \frac{\phi_c}{\kappa \sigma^2\gamma} 
\ee
Of course, this is so sensitive it would be immediately ruled out (by producing too many PBHs), but we can then decrease $N_f$ to obtain a viable window.   To see if this copious production and hence very strong sensitivity indeed arises, we next analyze the CMB constraint.  

Since we are working with $p=2$, we have an easily calculable spectrum of $\Lambda_n$ from (\ref{StarSchrod}).  The result is
\be\label{Lambdanp2}
\Lambda_n = \frac{\lambda_2}{2\pi ^2}n, ~~~~~~~~~~~~~~ p=2
\ee    

{

In general, the PBH mass (\ref{Mass}) is independent of the other parameters in our model, since it depends strongly on the time $t_{mix}$ at which we introduce the mixing interaction.  In particular, it is independent of $\lambda_2$.  The latter enters into the CMB constraint, suppressing the effects of $\chi$ on that scale as follows.  \footnote{In the case that the the mixing is around reheating, then 
\be\label{Mp2}
M=\frac{M_P^2}{H} ~~~~~~~  t_{mix}\sim t_{reheating}
\ee
%The PBH mass is 
%\be\label{Mp2}
%M= e^{2\gamma_c H/\Lambda_1}\frac{M_{pl}^2}{H}=e^{4\pi^2\gamma_c/\lambda_2}\frac{M_{pl}^2}{H}
%\ee
For this to be between $10^{15}gr$ and the mass of the universe, we have a corresponding window of $H/M_P$:
\be\label{lambdawindow}
 20\log(10)<\log(M_{Pl}/H)<55\log(10)  ~~~~~ t_{mix}\sim t_{reheating}
\ee
}

The two point function (subtracting the disconnected piece) 
is proportional to $N_f$, so the CMB constraint becomes
\be\label{CMBNf}
\frac{\kappa^2}{H^4} G(R_{CMB})\sim N_f\kappa^2\frac{1}{(R_{CMB}H)^{2\Lambda_2/H}}\sim N_f \kappa^2 e^{-2 N_e\lambda_2/\pi^2} < \frac{10^{-3}}{H^2}
\ee
In this example, from an exact calculation of the $\langle\chi^2\chi^2\rangle$ two point function, we find an order one coefficient which we will not keep track of here.  

So for $\beta\simeq 1$ we have two constraints on $N_f$, at fixed values of the other parameters:
\be\label{Nfwindow}
\frac{\phi_c}{\kappa\sigma^2\gamma} < N_f < \frac{10^{-3}}{\kappa^2 H^2}e^{2 N_e\lambda_2/\pi^2}
\ee
This is straightforward to satisfy on both ends.   

To spell this out, let us use the relations $\kappa = \Delta t \lambda_{mix}=\Delta t\dot\phi/M_*^2$, $\sigma=H/2\pi$, $\phi_c/\sigma\sim \zeta_c/\zeta_G$, $\zeta_c\sim 1/10$ along with the observed power spectrum amplitude $\zeta_G\sim 5.4\times 10^{-5}$ to write the lower bound on $N_f$ (for $\beta\simeq 1$) as
\be\label{lmixform}
\frac{\phi_c}{\kappa\sigma^2\gamma} =\frac{\zeta_c}{\zeta_G} \frac{2\pi}{(H\Delta t)\lambda_{mix}\gamma}=\frac{10^{-1}}{5.4\times 10^{-5}} \frac{2\pi}{(H\Delta t)\lambda_{mix}\gamma}\simeq \frac{10^4}{(H\Delta t)\lambda_{mix}\gamma}\gtrsim \frac{10^4}{\epsilon_{mix}(H\Delta t) (4\pi/N_f^{1/4})\gamma}
\ee 
where in the last inequality we used the requirement (\ref{lambdaNf}) for radiative stability, and defined $\epsilon_\lambda=\lambda_{mix}N_f^{1/4}/4\pi$.  This implies a lower bound on $N_f$ (for $\beta\simeq 1$) of
\be\label{Nflowertext}
N_f> \left(\frac{10^4}{\epsilon_{mix}4\pi\gamma (H\Delta t)}\right)^{4/3}
\ee
%For $\gamma\simeq 1$ and $H\Delta t\simeq 10$ this gives a lower bound on $N_f$ of $\sim 300/\epsilon_{mix}^{4/3}%for $\beta\simeq 1$.   
Similarly (again using (\ref{lambdaNf}) we can write the upper bound on $N_f$ from (\ref{Nfwindow}) as
\be\label{Nfupperagain}
N_f< \left(\frac{10^{-3}}{\epsilon_{mix}^2(4\pi)^2(H\Delta t)^2} e^{2 N_e \lambda_2/\pi^2}\right)^2
\ee  

Let us illustrate this with some numbers, involving a rather large $N_f$, and then comment on this regime as well as on the possibility of reducing $N_f$.
Plugging in $H\Delta t=4, N_e=60, \lambda_2=3/2, \epsilon_{mix}\sim 1/5$ yields a condition 
$ 2\times 10^4 < N_f < 7\times 10^5$.  
So far in this section, we have focused on the regime with an enormous abundance of PBHs, $\beta\simeq 1$, way above the basic requirement (\ref{basicbetaconstraint}).  This illustrates that a multifield landscape with a sufficiently large number of fields can be immediately constrained by PBH production.   In string theory, for example, there are numerous fields (including $2^D$ Ramond-Ramond fields which descend to axions), and it would be interesting to constrain string-theoretic models using this effect.     We could reduce $N_f$ and obtain a viable level of PBH production, although this is a rather sharp transition.  If, however, we had considered potentials with an outward drift in the $\chi$ sector, the non-Gaussian tail would be stronger, likely reducing $N_f$.  Finally, in the next section we will comment on an exponential form of the function $F(\chi)$ that enters into the mixing interaction, motivated by the structure of fields and couplings in string theory including hyperbolic field space geometries.   Regardless, even with a large $N_f$, the calculable non-Gaussian structure of the tails of our distribution is novel.

\section{Case of an exponential mixing interaction and super-massive black hole seeds}

%{\color{blue} ES: while everyone is checking the rest, we'll add a few comments about the hyperbolic case here, as well as adding references etc.  The remaining issue in the hyperbolic case was the radiative corrections which would be controlled by $V'$.  We expect not to have to tune that as much as in a Gaussian model.  Regardless, even if we needed to accidentally have $V'=0$ (e.g. a saddle point) during $\Delta t$, the structure of the tail is so novel here that it is worthwhile to point out this example in my view.  It, along with our freedom to mix earlier, gives us a mechanism for SMBH seeds.}

It is interesting to consider more general forms of $F(\chi)$ in (\ref{mixgen}) and more general Lagrangians during the mixing period.  One interesting possibility is a hyperbolic field space coupled to the inflaton  
during a mixing interval $\Delta t$, with action
\be\label{hyperbolictext}
\int \left\{ (\partial\chi)^2+e^{2\chi/M_*}(\partial\phi)^2-V(\phi)  \right\}
\ee
This involves an exponential of one of the fields (e.g. $\chi$ in our scenario); for recent work on inflation in this field space geometry see e.g. \cite{AndreiRenata}\cite{Brown:2017osf}.  As we discuss in the appendix the isometries of the space strongly limit the corrections in the limit $V'\to 0$ \cite{Brown:2017osf}.  Moreover, the exponential dependence on $\chi$ leads to a stronger tail.  In the appendix, we explore this example, finding a tail which behaves like
\be\label{tailhyp}
\rho(\delta\phi_0)\sim e^{-(\log\delta\phi_c^2)^2M_*^2/H^2}
\ee
as a function of the dimensionful parameter $M_*$ in the model.  
The shape of this tail also satisfies the criterion for super-massive black hole seed formation presented in \cite{SMBHmu}.  It would be interesting to derive the predictions and constraints on this model in detail, something we leave for future work.

\bigskip

\noindent{\bf Acknowledgements}
We thank Mehrdad Mirbabayi, Leonardo Senatore, and Matias Zaldarriaga for extensive discussions and collaboration on this subject, and D. Spergel for suggesting that we investigate the PBH application of our strongly non-Gaussian tails.    We are also very grateful to J.R. Bond and V. Gorbenko for discussions of the stochastic dynamics of inflationary perturbations, A. Brown for discussions of radiative corrections on hyperbolic field space, and to A. Linde, P. Michelson and C. Unal for interesting discussions of PBHs.   E.S. thanks the organizers and participants of the Princeton Center for Theoretical Science workshop `The Accretion Signatures of the Earliest Black Holes in the Universe'.    This research was supported in part by the Simons Foundation Origins of the Universe Initiative (modern inflationary cosmology collaboration), and by a Simons Investigator award.   GP and ES are grateful to the KITP for hospitality during part of this project.

\newpage

\appendix

{
%\color{magenta}

\section{Radiative stability of the result}

Here we analyze the path integral for the probability $\rho(\delta\phi_0)$ during the time interval $\Delta t$ between the mixing time $t_m$ and $t_0=t_m+\Delta t$, for a quadratic mixing interaction $F(\vec\chi)=\vec\chi^2/M_*^2$.   We will analyze it in two ways in the following two subsections.  In the first analysis, which will be our main check of radiative stability, we will integrate the short modes of a large number of $\chi$ flavors out first to obtain the effective action for the $\delta\phi$ field.   In combination with the analysis in the main text, we will exhibit radiative stability of our results.  Then in the second analysis, we will explore the possibility of integrating first over $\delta\phi$, which gives an interesting structure worthy of further study.  

Schematically, in our model the path integral which computes the probability for $\delta\phi_0$ is:
\bea\label{fullPI}
\rho(\delta\phi_0)&=& \int D\chi_0 \Psi^\dagger(\chi_0, \delta\phi_0) \Psi(\chi_0,\delta\phi_0)\nonumber\\ 
&=&\int D\chi_0 \int D\chi_m D\tilde\chi_m d\delta\phi_m d\delta\tilde\phi_m \int D\chi|_{b.c.} D\tilde\chi|_{b.c.} D\delta\phi|_{b.c.} D\tilde\delta\phi|_{b.c.}~\nonumber\\ & & ~~~~~~~~~~~~ e^{i (S -\tilde S)}\Psi_G(\delta\phi_m) \Psi_G(\chi_m)\Psi_G^\dagger(\tilde\chi_m)\Psi_G^\dagger(\delta\tilde\phi_m)  \nonumber\\
\eea
with boundary conditions $\chi(t_m)=\chi_m, ~ \chi(t_0)=\chi_0=\tilde\chi(t_0), ~ \tilde\chi(t_m)=\tilde\chi_m$ and similarly for $\delta\phi$.  
The action is
\be\label{actiondelt}
S=\int_{t_m}^{t_m+\Delta t} dt d^3\x \sqrt{-g} \left\{(\partial\chi)^2 + (\partial\delta\phi)^2 + \lambda_m \chi^2\delta\dot\phi +\dots\right\}, ~~~~ \lambda_{mix} = \frac{\dot\phi}{M_*^2}\equiv \lambda
\ee
where the $\dots$ refers to other interactions that may arise during $\Delta t$; specific examples are given in the following section.  
Here we note that $\lambda$ is not to be confused with the $\lambda_p$ in the main text, this is the mixing interaction.
and similarly for the tilded variables.  This path integral is completely Gaussian in $\chi$ and $\tilde\chi$.  We will derive the 1PI effective action as an analogue of the Coleman-Weinberg potential, although in our model $\chi^2$ couples to the derivative $\delta\dot\phi$ rather than $\delta\phi$ itself.  

\subsection{Integrating over the $\chi$'s}

In this subsection we will work with the interaction $\dot\phi^2\frac{\chi^2}{M_*^2}$, whose large cross term yields the mixing interaction in (\ref{actiondelt}).   We will also consider $N_f\gg 1$ flavors because that ensures that the 1-loop contribution to the effective action for $\delta\phi$ dominates,  provided 
\be\label{lambdaNf}
\lambda \equiv \frac{\epsilon_{mix} 4\pi}{N_f^{1/4}} < \frac{4\pi }{N_f^{1/4}}
\ee
This inequality will hold in our model, consistently with the large excursion on the tail scaling like $\lambda \vec\chi_{tail}^2\Delta t\sim \delta\phi_c$.  The 1-loop 1PI effective action obtained by integrating out the $N_f$ $\chi$ fields (inside the horizon) is given by 
\bea\label{actionfull}
\Gamma_{1PI}&=&\int a^3\left\{\frac{1}{2}\dot\phi^2\left(1-\frac{\vec\chi^2}{M_*^2}\right)+\frac{1}{2}(\partial\chi)^2-V(\phi) - \Delta V_{eff}(\chi)\right.\nonumber\\ 
&-&\left.\frac{N_f}{2}\int_H^{M_{UV}} \frac{d^4 k_E}{(2\pi)^4}\log\left(1+\frac{\dot\phi^2/M_*^2}{k_E^2-i\epsilon}\right)  \right\}\nonumber\\
\eea
This is the standard trace log from the determinant from integrating out the Gaussian $\vec\chi$ field.   By writing the $\Delta V_{eff}(\chi)$ we have allowed for new contributions to the $\chi$ effective action during the interval $\Delta t$.  
Here we dropped $\phi$ gradients for simplicity because they are long modes; the large $N_f$ expansion ensures that their contributions to loops is suppressed.
Below we will ensure that our tail falls in the regime
\be\label{regime}
\frac{\vec\chi_t^2}{M_*^2} <1 
\ee
so that the action (\ref{actionfull}) makes sense.  

The first question is whether the background solution is affected by the log term.  This is an interesting possibility in itself, in this regime where we have an expression with a resummed dependence on $\dot\phi$.   In this respect, our model is similar to DBI inflation and we leave an analysis of its dynamics for future work.    For now we will impose a condition on our parameters to avoid a dominant effect of the log term on the background solution.  By varying this effective action, we find that this is achieved in the regime
\be\label{bgok}
N_f M_{UV}^4 < {\dot{\bar\phi}}^2
\ee
and we will satisfy this for simplicity.  Note that here, the $\dot{\bar\phi}$ is the evolution during the $\Delta t$ mixing period (not necessarily the same as the overall inflationary slow roll value related to $\zeta$ and $\delta\phi_c$).  

Let us now expand the log term about the background solution:  
\be\label{logexpansion}
N_f\int_k \log(\dots) = const + N_f\int_k \log\left(1+\frac{2\dot{\bar\phi}\delta\dot\phi/M_*^2}{k_E^2+\dot{\bar\phi}^2/M_*^2} \right) = const + N_f\int_k \log\left(1+\frac{2\lambda\delta\dot\phi}{k_E^2+\lambda\dot{\bar\phi}} \right)
\ee
where the constant term here (which does not matter for the dynamics) is $N_f\int_k \log(1+(\dot{\bar\phi}/(M_*^2 k_E^2)))$, and we used that $\delta\dot\phi\ll \dot{\bar\phi}$.   

The second term inside the log is $\ll 1$ since $\delta\dot\phi\ll\dot\phi$.  The coefficient of the log term is large, so let us expand it to see if its leading terms compete with the existing ones. 
After integrating by parts, the linear term is of order 
\be\label{linearterm}
-\delta\phi ~ \lambda N_f \int_H^{M_{UV}}\frac{d^4 k}{(2\pi)^4}\frac{1}{a(t)^3}\frac{d}{dt}\left(\frac{a(t)^3}{k^2+\lambda\dot{\bar\phi}}\right)
\ee
As mentioned above, this is subdominant in its contribution to the equation of motion compared to the leading terms of order $3H\dot\phi\delta\phi$ for the background evolution, provided we satisfy (\ref{bgok}).  

Next consider the radiative correction to the quadratic term:
\be\label{quaddot}
\delta\dot\phi^2 N_f \lambda^2 \int_H^{M_{UV}} \frac{d^4 k}{(2\pi)^4}\frac{1}{(k^2+\lambda\dot{\bar\phi})^2}
\ee
%This is essentially the effect Leonardo computed earlier, with one difference/generalization:   
In the denominator here, we see the effective mass squared of $\chi$ from the background.
%(Leonardo worked in the EFT of perturbations and did not have (or need) this effect.)  
This effective mass term can suppress the radiative effect, but there is a trade-off:  
it also causes the $\chi$ modes to find themselves on
an effective potential hill, with $m_{eff}^2\sim \lambda \dot\phi$, pushing them to smaller values.  One could restrict $\Delta t$ to a small enough timescale to avoid a significant decrease of $|\vec\chi|$ during this period.
Instead, we will accompany the mixing with a linear term in $\Delta V(\chi)$ which pushes outward on $\chi$ to compensate this.
This can be done with reasonable scales in the potential: near our tail excursion of $\vec\chi$ we need 
\be\label{Vcomp}
\Delta V + \lambda\dot\phi\vec\chi^2 \sim -\mu^3 |\vec\chi| + \frac{1}{2}\lambda\dot\phi\vec\chi^2
\ee
with 
\be\label{muscale}
\mu^3 \sim \lambda\dot\phi |\vec\chi_{tail}|\sim \dot\phi \sqrt{\frac{\lambda\delta\phi_c}{\Delta t}}\sim\dot\phi H \sqrt{\frac{4\pi \epsilon_{mix}\phi_c}{N_f^{1/4}\Delta t H^2}}
\ee

%Now for $\beta\simeq 1$ we use (\ref{ysaddlecond}) and (\ref{lmixform}) and the relation $\kappa=\lambda_{mix}\Delta t$, we 
%have {\color{blue} update with $\pi$'s:}
%\be\label{Deltlowerbd}
%\frac{N_e}{H}> \Delta t > \frac{\delta\phi_c}{\lambda_{mix}N_f H^2} \Rightarrow \dot\phi \sqrt{\frac{\lambda\delta\phi_c H}{N_e}}< \mu^3 < H\dot\phi \lambda\sqrt{N_f} 
%\ee 
%This is a reaonable window of scales for $\mu^3$. It also imposes a lower bound on the number of flavors $N_f$:
%\be\label{Nflower}
%N_f > \left(\frac{\delta\phi_c}{4\pi H N_e} \right)^{4/3}
%\ee    
%where we used (\ref{lambdaNf}).  This is a large number of flavors for $\beta\simeq 1$, but we note that some scenarios for UV complete physics such as string theory contain numerous scalar fields and we may use this effect to exclude them empirically (given the appropriate couplings).  This relation (\ref{Nflower}) pertains to an overabundance $\beta\simeq 1$ since it assumed (\ref{ysaddlecond}), and $N_f$ can be lowered somewhat from the value on the RHS of (\ref{Nflower}) to produce a viable abundance.     
%%(e.g. $2^D$ Ramond-Ramond fields which descend to axions)     

%However, we can proceed regardless.  In the absence of other contributions to the $\chi$ dynamics during $\Delta t$, to keep the $\chi$ modes (e.g. those interest for our tail) in place, this requires us to limit $\Delta t$ to be less than $1/m_{eff}$.  

We find a mild lower bound on $\Delta t$ from the requirement that $\frac{\vec\chi_{tail}^2}{M_*^2}=\frac{\delta\phi_c}{\lambda \Delta t M_*^2}=\frac{\delta\phi_c}{\dot\phi}<1$; this corresponds to $\Delta t>\frac{\delta\phi_c}{\dot\phi}$.   Given that $\dot\phi\ge 10^5 H^2$ (taking it to be at least as large as the overall inflationary $\dot\phi$), this requirement is satisfied in our regime, e.g. for the value $H\Delta t=4$ used as an example in the main text.

%Altogether we can demand a window
%\be\label{Deltatwindow}
%\frac{\zeta_c}{H}\frac{\dot\phi_{inf}}{\dot\phi}< \Delta t < \frac{N_e}{H}
%\ee  
%where $\dot\phi_{inf}$ is the value of $\dot\phi$ overall in inflation (while our $\dot\phi$ during $\Delta t$ may be larger, e.g. if its potential steepens during this period).  Also, this upper bound is not needed if the potential for $\chi$ develops a linear piece which pushes it outward during $\Delta t$.  In general, both this drift of $\chi$ and the size of $\dot\phi$ during the $\Delta t$ period are model dependent and amenable to relatively simple model building.

We have other inequalities as follows.  
In order for the radiative effect to be suppressed we want $\lambda\dot\phi>M_{UV}^2$, along with the above condition (\ref{bgok}).  We also want $M_{UV}\gg H$ so that we are not tuning the effective UV cutoff of the $\chi$ loops.  So we want
\bea\label{ineqs}
H^2 \ll M_{UV}^2 \ll \lambda\dot\phi ~~~and ~~~~H^2 \ll M_{UV}^2 \ll  \frac{\dot\phi}{\sqrt{N_f}}
\eea 
which just requires $\lambda \gg H^2/\dot\phi, N_f\ll \dot\phi^2/H^4$, which are both easily satisfied.  With $\lambda\sim 1/N_f^{1/4}$, the first of these inequalities is the stronger one.  

With these windows in place, which are straightforward to satisfy, let us go back to the radiative correction (\ref{quaddot}).  This is of order 
\be\label{radcorrquad}
\delta\dot\phi^2 N_f\lambda^2 \frac{M_{UV}^4}{(\lambda\dot\phi)^2}=\frac{N_f M_{UV}^4}{\dot\phi^2}\ll 1 
\ee
and can be neglected compared to the classical term.

%Again, this suppression of the radiative correction looks consistent with the tail of $\vec\chi$ cancelling the Gaussian suppression at $\delta\phi=\delta\phi_c$.  With $\Delta t$ bounded above, $\chi$ will not
%roll down from the tail.  (Again if it had we could let it drift outward during $\Delta t$ instead, with a linear term in its potential.)  

%Given that, we can write the 1PI effective action for the perturbation $\delta\phi$:
%\bea\label{1PIdelt}
%\Gamma_{1PI}(\delta\phi)&=&\int a^3\{\frac{1}{2}(\dot\delta\phi)^2(1-\frac{\vec\chi^2}{M_*^2})
%-\lambda\delta\dot\phi \vec\chi^2
%+\frac{1}{2}(\partial\chi)^2\nonumber\\ 
%&-&\frac{N_f}{2}\int_H^{M_{UV}} \frac{d^4 k_E}{(2\pi)^4}\log(1+\frac{(\dot{\bar\phi}+\delta\dot\phi)^2/M_*^2}{k_E^2-i\epsilon})  \}\nonumber\\
%\eea
%where (again for simplicity of the expression) we took $V$ linear in $\delta\phi$, and we used the fact that the log term's contribution to the equations of motion, i.e. the linear term in $\delta\phi$, is subdominant (we can subtract a $\delta\log(\dots)$ piece but that was negligible).  

%Ok now we are ready to compute the field momentum and determine the implication of the radiative correction:
%\be\label{Pifull}
%\frac{\Pi_{\delta\phi}}{a^3}\simeq \delta\dot\phi(1-\frac{\vec\chi^2}{M_*^2})-\lambda\vec\chi^2-\frac{N_f M_{UV}^4}{4 \dot\phi} \simeq \delta\dot\phi-\lambda\vec\chi^2
%\ee
%where the last expression follows from the regime we specified above.  

\subsection{Comment on another order of integration}

We can also work with the path integral in another order of integration, starting with $\delta\phi$.  
In this subsection, we comment briefly on this approach, which leads to an interesting structure.  However, it is unconventional in the sense that the long modes of $\delta\phi$ are integrated out before the short modes of $\chi$, leading to a non-local expression.  However, as we will see, this expression is very intuitive from the perspective of our mechanism for strong non-Gaussian tails.  

\be\label{psiPI}
 \Psi(\chi_0, \delta\phi_0) = \int D\chi_m D\phi_m \int D\chi|_{b.c.} D\delta\phi|_{b.c.} e^{i S} \Psi_G(\delta\phi_m)\Psi_G(\chi_m)
  \ee
  We can make the following shift in $\delta\phi$:
  \be\label{dphishift}
  \delta\phi_k\rightarrow\delta\phi_k-\frac{1}{2} \lambda \int_{t_m}^t dt^{\prime} (\chi^2)_k(t^{\prime})
  \ee
  Then the action becomes:
  \bea\label{actionshifted}
  S &=& \int_{t_m}^{t_m+\Delta t}dt a(t)^3 d^3\k \left\{\delta\dot\phi_{\k}\delta\dot\phi_{-\k}-\frac{k^2}{a(t)^2}\delta\phi_\k\delta\phi_{-\k} \right.\nonumber\\ 
  &+&  \left.\dot\chi_\k\dot\chi_{-\k}  - \frac{1}{4} \lambda^2 (\chi^2)_\k(\chi^2)_{-\k}  
  +\lambda^2 \frac{k^2}{a^2}\left(\int^t_{t_m}dt^{\prime} \chi(t^{\prime})^2_\k\right)\left(\int^t_{t_m}dt^{\prime} \chi(t^{\prime})^2_{-\k}\right)   \right\}
  \eea
  while the boundary conditions are $\delta\phi_m\rightarrow \delta\phi_m$ and for each $k$ mode
  \be\label{dphi0shift}
  \delta\phi_0\rightarrow \delta\phi_0^{\prime} = \delta\phi_0 - \frac{1}{2} \lambda \int_{t_m}^{t_m+\Delta t} dt \chi^2
  \ee
  In the action, we have neglected the term:
  \be\label{shiftednegl}
  2\lambda \partial^i \delta\phi \int_{t_m}^t dt^{\prime} \chi(t^{\prime})\partial_i\chi(t^{\prime})
  \ee
  This term is suppressed by a factor of the field gradients as well as a factor of $\lambda$ and $\Delta t$.

  The idea behind this shift is to decouple $\delta\phi$ and $\chi$ in the action (they are still coupled via the boundary condition $\delta\phi_0^{\prime}$). Now, the $\phi$ integrals of (\ref{psiPI}) are just the evolution of a Gaussian state in DS and give another Gaussian state $\Psi_{G,evolved}(\delta\phi_0^{\prime})$. Then,
  \be\label{psiPIshifted}
  \Psi(\chi_0, \delta\phi_0) = \int D\chi_m  D\chi|_{b.c.}  e^{i S_{\chi}} \Psi_G(\chi_m)\Psi_{G,evolved}\left(\delta\phi_0 - \frac{1}{2} \lambda \int_{t_m}^{t_m+\Delta t} dt \chi^2\right)
  \ee
  where $S_{\chi}$ is the action (\ref{actionshifted}) without the first term. This resembles (\ref{Psiafter}). The next step would be to find a saddle solution for $\rho(\delta\phi_0)$.
  
Within the expression for $\rho(\delta\phi_0)$ (\ref{fullPI}), there is a saddle point satisfying $\chi_s=\tilde\chi_s^*$.  In this expression, the contribution from the inside-horizon modes of $\chi$ are helpful in cancelling the tail in $\delta\phi$.  It appsears from this expression  $\chi$ trajectories can evolve from $\chi_m$ to $\chi_0$ with the integral in (\ref{psiPIshifted}) cancelling the tail.  It would be interesting to pursue this analysis further.  However, the integration performed in the other order in the previous subsection suffices to establish radiative stability.

%\section{Generalities}

\section{Hyperbolic field space case as a candidate for a large tail with radiative stability}

%{\color{blue}  ES:  While you are checking the $\vec\chi^2$ model, we'll see what can be done with this model and import the conclusions or conjectures about it to the final section in the main text.}

Consider the theory during our mixing interval $\Delta t$ with action
\be\label{hyperbolic}
\int \left\{ (\partial\chi)^2+e^{2\chi/M_*}(\partial\phi)^2-V(\phi)  \right\}
\ee
The metric part of this has the continuous isometries of hyperbolic space (as in certain supergravity models \cite{AndreiRenata}\ and in the recent inflationary mechanism \cite{Brown:2017osf}).  The isometry group includes a continuous shift in $\chi$ combined with a rescaling of $\phi$.  Up to effects suppressed by $V'$, the only correction to the two derivative kinetic term that is possible is
to change $M_*$; let us denote by $M_*$ the final value of it.  In fact, as argued in \cite{Brown:2017osf}, a small value of $M_*$ is well-motivated analogously to small axion decay constants.    Higher derivative terms may arise, but these must respect the isometries.   For example, powers of $(\partial\phi)^2$ must arise in the combination $e^{2\chi/M_*}(\partial\phi)^2$, and below we will see that $e^{2\chi/M_*}$ will be small in our regime of interest.  Also, both for these corrections and higher powers of $(\partial\chi)^2$, we have suppression by powers of $1/a(t)$. 
The other radiative corrections vanish in the limit that $V'\to 0$.  A value of $V'$ which is exponentially suppressed as a function of inverse coupling constants may be obtained in a Wilsonian natural way via dimensional transmutation.  In what follows, we assume that $V'$ is small enough for the leading effects in the mixing period $\Delta t$ to come from the first two terms in (\ref{hyperbolic}).  

Let us define $f(\chi)=e^{2\chi/M_*}$ (note that this is not the same as $F(\chi)$ in the main text).
During our small interval of time $\Delta t$ for mixing, we get a relatively simple evolution operator $U$:
\be\label{Udelt}
\Psi(t_0) \simeq U \Psi(t_0-\Delta t)\sim e^{-i\Delta t \int_x \frac{\hat\Pi_\phi^2}{a(t_0)^3 f(\chi)}} \psi_G(\phi)\psi_G(\chi)
\ee
We note that in our book-keeping here, we will include the whole field $\phi$ (the approximately frozen background and the perturbations).  
The exponent  is not just a shift of $\phi$, but as we will see below it includes that and it is still very tractable since it is Gaussian in the $\phi$ sector.  (The ultimate distribution will be highly non-Gaussian from the $\chi$ integral.)  In writing just the $\hat \Pi_\phi^2$ term in the Hamiltonian, we used the fact that the other terms acting on our state will be small in the small-$F$ regime we will be interested in.  Note that here unlike before we are not dropping the $\delta\dot\phi^2$ term from the start, instead we keep the full interaction between $\chi$ and $\phi$.  

Let us write it in a field momentum basis in order to compute the effect of this step of evolution.  We use a matrix notation for the position space integrals.  
\be\label{psiGFT}
\psi_G(\phi)=e^{-\frac{1}{2}(\phi-\bar\phi) \Sigma^{-1}(\phi-\bar\phi)}e^{i{\bar{\Pi}}_\phi\phi}=\int D\Pi_\phi e^{-\frac{1}{2}(\Pi_\phi-\bar\Pi\phi) \Sigma (\Pi_\phi-\bar\Pi_\phi)} e^{i \Pi_\phi (\phi-\bar\phi)}
\ee
with 
\be\label{background}
{\bar{\Pi}}_\phi =a_0^3 ~{\overline{f \dot{\phi}}}
\ee
This yields
\bea\label{psievolved}
\Psi(t_0) &\simeq& \psi_G(\chi_0)  \int d\Pi e^{i\Pi\cdot (\phi-\bar\phi)}e^{-\frac{1}{2}(\Pi-\bar\Pi\phi) \Sigma (\Pi-\bar\Pi_\phi)}e^{-i\Delta t \Pi^2/a_0^2 f(\chi)}  \nonumber\\ 
&=& phase\times \psi_G(\chi_0)  \nonumber\\
&& \exp\left(-\frac{1}{2}\bar\Pi_\phi \Sigma\bar\Pi_\phi-\frac{1}{2}[\phi-\bar\phi-i{\bar\Pi_\phi}\Sigma] \left(\Sigma+i\frac{\Delta t}{a_0^3 f(\chi_0)} \right)^{-1}[\phi-\bar\phi-i\Sigma\bar\Pi_\phi]\right)\nonumber\\
\eea
There are two effects here:  a shift in field space and a wider width.  We will unpack the formula to exhibit each of these next. 

First, consider expanding in the $\Delta t$ term.  Let us simplify the notation by writing the matrix
\be\label{Deltadef}
\Delta_{T x, x'}\equiv \frac{\Delta t}{a_0^3 f(\chi_0(x))}\delta_{x, x'}
\ee
This is diagonal (but not proportional to the identity).   We have
\be\label{inverseexp}
(\Sigma+i\Delta_T)^{-1} = \Sigma^{-1}-i\Sigma^{-1}\Delta_T\Sigma^{-1}+{\cal O}(\Delta_T^2)
\ee 
This gives for (\ref{psievolved})
\be\label{psiexpand}
phase\times \exp\left(-\frac{1}{2}[\phi-\bar\phi]\Sigma^{-1}[\phi-\bar\phi]+ \bar\Pi_\phi \Delta_T \Sigma^{-1}[\phi-\bar\phi]+{\cal O}(\Delta_T^2)\right)
\ee
So at this order in the exponent, we see the shift 
\be\label{shiftphi}
\phi-\bar\phi\to \phi-\bar\phi -\Delta_T\bar\Pi_\phi =\phi-\bar\phi-\Delta t\dot{\bar\phi}-\Delta t\left(\frac{\overline{f\dot\phi}}{f}-\dot{\bar\phi}\right)=\delta\phi--\Delta t\left(\frac{\overline{f\dot\phi}}{f}-\dot{\bar\phi}\right)
\ee
If we write $f=1+F$ and perturb in $F$ this is
\be\label{Fpert}
\delta\phi -\Delta t(\overline{F \dot\phi}-F\bar{\dot\phi}) ~~~~ F\equiv 1-f\to 0
\ee
which is similar to the examples in the main text.  

%we see an analogue of our familiar shift in field space (although note that before, $F$ was the deviation from the Euclidean metric on field space where as here $f$ multiplies the whole $\phi$ kinetic term).   

We also see a different regime where $f(\chi)=e^{2\chi/M_*}<<1$, the $\Delta t$ term dominates, and the width for $\phi-\bar\phi$ is enormous.   We pay a small price for this in $\psi_G(\chi)$ because $\chi\sim M_*\log f$.  
The calculation of the probability distribution for $\delta\phi$ proceeds as before, $\rho(\delta\phi)=\int D\chi_0 |\Psi|^2$.   A toy one-dimensional analogue of the $\chi_0$ integral behaves as
\be\label{chi1d}
\int dx e^{-x^2/2} e^{-(\delta\phi_c)^2/(1+ e^{-4x})}
\ee
In other words, even ignoring the field space shift, the saddle is at $4x \approx -\log\delta\phi_c^2$, and the suppression is only by $e^{-(\log\delta\phi_c^2)^2M_*^2/H^2}$, giving us a highly non-Gaussian tail.  Recall that $M_*$ here can naturally be small, corresponding to a highly curved hypebolic metric.  

%This has a simple intuitive explanation: the exponential $f\sim e^{2\chi/M_*}$ is similar to what we would get by taking our previous parameter $m$ to infinity in $f\sim \chi^m$.  This exponential is chosen here for naturalness since it gives us a maximally symmetric (hyperbolic) field space metric.  

\section{Analytic results for correlations and distributions in the $p=1$ example}

In this appendix we would like to collect some formulas from \cite{Starobinsky}\ for the equilibrium two point function, and then evaluate it for the $p=1$ model in the $\chi$ sector.    We will then generalize to the two point function of $\chi^2$ as in \cite{GS}.   Finally, we will analyze the joint probability distribution.  In this appendix, we will for simplicity work with a single $\chi$ field; the generalization to $N_f>1$ is straightforward and is incorporated in the main text.  

Following the notation in \cite{Starobinsky}, but for simplicity setting $H=1$ in this subsection, we have an equal-time two point correlation function (cf \cite{Starobinsky} eqn (84)):
\be\label{Grtt}
G(R) = N\sum_n |A_n|^2 e^{-2\log(R)\Lambda_n} 
\ee
where 
\be\label{Andef}
A_n = N^{-1}\int d\chi \chi e^{-\lambda_p |\chi|^p}\Phi_n(\chi)
\ee
with
\be\label{Ndef}
N=\int_{-\infty}^\infty d\chi e^{-2\lambda_p |\chi|^p} = \frac{2^{1-1/p}}{\lambda_p^{1/p}}\Gamma\left(1+\frac{1}{p}\right) 
\ee

We would like to evaluate this for $p=1$, where the spectrum of (\ref{StarSchrod}) has a continuum above a gap, with  simple eigenfunctions 
%{\color{blue} See below for more explicit normalized wavefunctions with the right boundary conditions at the delta function at the origin}
\be
\Phi_{\pi_\chi}\sim e^{i\pi_\chi \chi},
\ee
and corresponding eigenvalues
\be\label{ponePhiLambda}
\frac{\Lambda}{H}=\frac{H^2}{8\pi^2}\left(\pi_\chi^2+\frac{\lambda_1^2}{H^2}\right).
\ee
To be precise, we should include the correct boundary conditions at the origin.
We find two wavefunctions (one odd and one even):
\be\label{odd} 
\Phi_{odd}(\chi) = \frac{1}{\sqrt{\pi}}\sin(\pi_{\chi} \chi) 
\ee
and
\be\label{even}
\Phi_{even}(\chi) = \frac{1}{\sqrt{\pi}}\sin(\pi_{\chi} |\chi| - \arctan(\pi_{\chi} H/\lambda_1)) 
\ee
The normalization constant is chosen so that:
\[ \int d\chi \Phi_{\pi_1}(\chi) \Phi_{\pi_2}(\chi) = \delta(\pi_1 - \pi_2) \]

If we include the full infinite $\chi$ field space, the sum over $n$ in (\ref{Grtt}) becomes a continuum integral over the field momentum $\pi_\chi\simeq \sqrt{\Lambda}$, which we will take into account in our calculations here.  Of course, physically only a finite range of $\chi$ is involved (for which the potential $V(\chi)$ remains below the inflationary scale).  In fact, the coefficients $A_n$ suppress the $\chi\to\infty$ regime, so it will self-consistently not contribute.

In this case, we find (again in $H=1$ units)
\bea\label{NAone}
N_{p=1} &=& \frac{1}{\lambda_1} \nonumber\\
%\Phi_{\pi_\chi} &=& e^{i\pi_\chi\chi}\nonumber\\
A_{\pi_\chi} &=& \frac{4 \pi_\chi/\lambda_1^2}{\sqrt{\pi}((\pi_\chi/\lambda_1)^2+1)^2}
\eea                                
which leads to  
\bea\label{GRone}
G(R)_{p=1} &=& \frac{1}{\lambda_1} \int d\pi_\chi \frac{|A_{\pi_\chi}|^2}{R^{2\Lambda}}\nonumber\\
&=&{\frac{1}{2\pi}} \frac{16}{\lambda_1^2}\frac{1}{R^{\lambda_1^2/4\pi^2}}\int_{-\infty}^\infty dw  \frac{w^2}{(1+w^2)^4}\exp\left(-\frac{\lambda_1^2}{4\pi^2}w^2\log(R)\right)
\eea
where to get the second expression we changed variables to $w=\pi_\chi/\lambda_1$.  
Here the factor of two downstairs is because we should only count $\pi_\chi>0$ since the others are redundant, and we define the integral over $w$ from $-\infty$ to $+\infty$.  
We note that the solutions $\Phi(\chi)$ which contribute are the odd ones $\sim \sin(\pi_\chi \chi)$.   The $R$-dependent prefactor is precisely (\ref{GRgen}), but as discussed below that equation in the main text, we in general have contributions of subleading non-equilibrium modes.  Here in (\ref{GRone}) this is reflected in the $R$-dependence in the integrand.   However, this has a mild effect:  a saddle point estimate gives $w_{saddle}^2\sim \frac{4\pi^2}{\lambda_1^2\log(R)}$, indicating a log rescaling.  As a result, our parametric estimate for the correlation length is not substantially affected, and we may use (\ref{GRgen}) as a reasonable estimate also for the $p=1$ example.   

{ 

As described in the main text, for our $m=2$ model we are also interested in the two point function  $\langle\chi^2(x^1)\chi^2(x_2)\rangle$ on CMB scales, to understand $G(R_{CMB})$ and apply the constraint that the second term in (\ref{twopoint}) not dominate.  This calculation is very similar to the calulation of $G(R)$ reviewed above, but now instead of $A_n$ defined above in (\ref{Andef}), we have an extra power of $\chi$ in the integrand.  Defining 
\be\label{Antwodef}
A_n^{(2)} \equiv N^{-1}\int d\chi \chi^2 e^{-\lambda_p |\chi|^p}\Phi_n(\chi),
\ee
the two point function of $\chi^2$ is given by  
\be\label{Gtwoortt}
G^{(2)}(R) = N\sum_n |A_n^{(2)}|^2 e^{-2\log(R)\Lambda_n} 
\ee
Now for an even potential $v(\chi)$ as we are considering, the eigenfunction $\Phi_1$ will be odd, and hence the leading contribution to $G^{(2)}$ will be $\Phi_2$.  This leads to the expression (\ref{GCMBexample}) in the main text.  For $p=1$, as we consider there, the spectrum of $\Phi_n$'s consists of a continuum above the gap at $\Lambda_1$.  So in that example, $\Lambda_2\simeq \Lambda_1 = \lambda_1^2/8\pi^2$, as in the last expression in (\ref{GCMBexample}).  

}

{

With the even wavefunctions given above, we find
\[ A_{\pi_{\chi}}^{(2)} = \frac{8\pi_{\chi} \lambda_1^2}{(\pi_{\chi}^2 + \lambda_1^2)^{5/2}\sqrt{\pi}} \]
Then (\ref{Gtwoortt}) becomes
\be\label{GCMBfull}
G^{(2)}(R)_{p=1} = \frac{32}{\pi \lambda_1^4}\frac{1}{R^{\lambda_1^2/4\pi^2}}\int_{-\infty}^{\infty} dw \frac{w^2}{(1+w^2)^5}\exp\left(-\frac{\lambda_1^2}{4\pi^2} w^2\log(R)\right)
\ee
Once again, this includes a factor of $1/2$ from the redundancy of considering both $\pm \pi_\chi$ for our even wavefunctions.  
In our analysis developed in the main text, we apply this at CMB scales, so $\log(R)=\log(R_{CMB})=N_e$.  

It is interesting to note that the analysis of the probability distributions $\rho_N(\chi_1,\dots, \chi_N)$ along the lines of \cite{Starobinsky}\ is particularly tractable in our $p=1$ model.  For example, the joint distribution $\rho_2$ is explicitly calculable as we will see in the following subsection.  

%We will now evaluate this for some of our examples in the main text.  Since the CMB condition was the most constraining, we saturate that.  That gives us the relation $120 \frac{\lambda_1^2}{8\pi^2}=\log(\kappa^2)$.  As a result, the prefactor of $\frac{1}{R^{\lambda_1^2/4\pi^2}}$ becomes
%\be\label{prefactorsat}
%\text{coefficient} = \frac{32}{\pi \lambda_1^4} \int_{-\infty}^{\infty} dw \frac{w^2}{(1+w^2)^5}\exp\left(-\log(\kappa^2) w^2\right)
%\ee
%We can numerically evaluate the integral in explicit examples.  
%We have a case in (\ref{CMBcondexp}) above with $\lambda_1=2$, $\kappa=30 \Rightarrow\log(\kappa^2)=6.8$.  Doing the integral numerically and putting in the factors outside yields
%that this coefficient is $.01$, i.e. we have a correction which is down by $10^{-2}$ from the inflaton contribution:
%\be\label{GCMBresult}
%\Delta G_{CMB} \sim 10^{-2} G_{CMB, usual} 
%\ee   
%This is not quite the $10^{-3}$ for the power spectrum amplitude, but as discussed above, it is not clear why we should impose that.  This $10^{-2}$ would likely translate to something that is enough to suppress the contribution to $f_{NL}$ (which would be constrained as $\kappa^3 1/R^{3 \Lambda/H}\times A^3 < 10^{-3}$.
%
%Another example is the supermassive BH case, which has $\kappa\sim 30, \lambda_1\sim \sqrt{3}$.  For this the coefficient is .03, which is again down by $10^{-2}$.  

}

{

{

\subsection{Joint probability}

%In the Starobinsky equilibrium configuration, a typical realization of the field fluctuates about $\chi=0$, with some rare large fluctuations on the tail.  We want to make sure we understood the spatial scale of these fluctuations.  
%
%In particular, to assess black hole production, we need to determine the spatial scale $R_{BH}$ on which
%the field evolves from $\zeta=\zeta_c$ to $\zeta=0$.  We note that requiring the field to evolve from $\zeta_c$ to 0, we avoid counting small spikes that only reach $\zeta_c$ because they lie on
%top of a broader overdensity.   In principle this scale could be anything between $H^{-1}$ and $R_{CMB}$, and we want to pay particular attention to whether it is simply the scale ``$R_c$'' that enters, or a shorter scale.  

In this section, we analyze the joint probability distribution for the field $\chi$ on two different patches.  We can use this, for example, to determine the spatial scale on which $\chi$ varies from a large value $\chi_t$ on its tail down to $\chi=0$.
Consider the joint probability $\rho_2(\chi_1(\x_1, t), \chi_2(\x_2, t))$, and define $a(t)|\x_1-\x_2|=R$.   We will be interested in how this behaves for $\chi_1(\x_1, t)=\chi_t, \chi(\x_2, t)=0$.    
      
This is easy to obtain (starting from \cite{Starobinsky} eqns (79-80)).  Doing the $\chi_r$ integral gives
\begin{equation}
\rho_2(\chi_1(x_1, t), \chi_2(x_2, t))=e^{-v(\chi_1)}e^{-v(\chi_2)}\frac{1}{N}\sum_{n=0}^\infty\Phi_n(\chi_1)\Phi_n(\chi_2)e^{-2\Lambda_n\Delta t}
\end{equation}
where 
\begin{equation}\label{Delt}
\Delta t=t-t_r = \log(RH\epsilon)
\end{equation}
in the notation of \cite{Starobinsky}.    
We will keep $\epsilon$ explicit; the field is smoothed on a scale $H^{-1}/\epsilon$.   We note that for $\Delta t=0$, the joint distribution reduces to 
\be\label{bc}
\rho_2\to \rho_1(\chi_1) \delta(\chi_1+\chi_2), ~~~~~ \Delta t\to 0
\ee
obeying the boundary condition noted in \cite{Starobinsky}.  

By separating out the $\Phi_0=\frac{1}{\sqrt{N}}e^{-v}$ contribution, we can write this in general as
\begin{equation}
\rho_2(\chi_1(x_1, t), \chi_2(x_2, t))=\rho_1(\chi_1)\rho_2(\chi_2)+e^{-v(\chi_1)}e^{-v(\chi_2)}\frac{1}{N}\sum_{n=1}^\infty\Phi_n(\chi_1)\Phi_n(\chi_2)e^{-2\Lambda_n\Delta t}\equiv \rho_1\rho_1+\Delta\rho_2
\end{equation}
When the first term dominates, the distribution decomposes into a product of the 1 point pdf's on each patch.
We can check when this happens as a function of $\chi_1, \chi_2$ and $RH\epsilon$.  In particular, we wish to understand the $R$ dependence when $\chi_1=\chi_t, \chi_2=0$ for some appropriate smoothing of the field determined by $\epsilon$. 

For our $p=1$ model, we can work the correction term out explicitly using the precise eigenvalues (\ref{ponePhiLambda}) and eigenfunctions (\ref{odd}-\ref{even}) above.  This is given by  (setting $H=1$ to simplify the formulas)
\be\label{oddevenform}
\Delta \rho_2 = e^{-v(\chi_1)}e^{-v(\chi_2)}\frac{1}{N}e^{-\lambda_1^2\Delta t/4\pi^2}
\frac{1}{2}\int_{-\infty}^\infty d\pi_\chi e^{-\pi_\chi^2\Delta t/4\pi^2}\left(\Phi_{odd}(\chi_1)\Phi_{odd}(\chi_2)+\Phi_{even}(\chi_1)\Phi_{even}(\chi_2)\right)
\ee
For our question as discussed above, we set $\chi_2=0$, and this simplifies the calculation since $\Phi_{odd}(0)=0$.  So it becomes 
\bea\label{nextrhodef}
\Delta \rho_2 &=& e^{-v(\chi_t)}e^{-v(0)}\frac{1}{2\pi N}e^{-\lambda_1^2\Delta t/4\pi^2} \nonumber\\
&&\int_{-\infty}^\infty d\pi_\chi e^{-\pi_\chi^2\Delta t/4\pi^2}\sin(\pi_{\chi} |\chi_t| - \arctan(\pi_{\chi}/\lambda_1)) \sin(- \arctan(\pi_{\chi}/\lambda_1)) 
\eea  
Next we simplify the integrand:  defining $w=\pi_\chi/\lambda_1$ 
\be\label{arcsimp}
\sin(\pi_{\chi} |\chi_t| - \arctan(\pi_{\chi} /\lambda_1)) \sin(- \arctan(\pi_{\chi} /\lambda_1)) =\frac{w^2\cos(w \lambda_1|\chi_t|)-w\sin(w \lambda_1|\chi_t|)}{w^2+1}
\ee
So we have
\be\label{rhodefw}
\Delta \rho_2 = e^{-v(\chi_t)}e^{-v(0)}\frac{\lambda_1}{2\pi N}e^{-\lambda_1^2\Delta t/4\pi^2}
\int_{-\infty}^\infty dw e^{-w^2\lambda_1^2\Delta t/4\pi^2}
\frac{w^2\cos(w \lambda_1|\chi_t|)-w\sin(w \lambda_1|\chi_t|)}{w^2+1}
\ee 
We can further use symmetry to write this in terms of phases as
\bea\label{rhodefwsymm}
\Delta \rho_2 &=& e^{-v(\chi_t)}e^{-v(0)}\frac{\lambda_1}{2\pi N}e^{-\frac{\lambda_1^2\Delta t}{4\pi^2}}
\int_{-\infty}^\infty dw ~ e^{-\frac{w^2\lambda_1^2\Delta t}{4\pi^2}}~
\frac{w^2e^{iw \lambda_1|\chi_t|}-(w/i) e^{i w \lambda_1|\chi_t|}}{w^2+1} \nonumber\\
&=&  e^{-v(\chi_t)}\frac{\lambda_1}{2\pi N}e^{-\frac{\lambda_1^2\Delta t}{4\pi^2}}
\int_{-\infty}^\infty dw ~ e^{-\frac{w^2\lambda_1^2\Delta t}{4\pi^2}}~e^{iw \lambda_1|\chi_t|}
\left(1+\frac{i}{w-i}\right)
\eea
Before working with this result, let us first check how it reduces to (\ref{bc}) for $\Delta t=0$.  The first term in parentheses gives a delta function, saturating the boundary condition (\ref{bc}).   The second term can be done by residues, with the integral over $w$ being simply $-2\pi  e^{-\lambda_1|\chi_t|}$.  At $\Delta t=0$, this second term in $\Delta\rho_2$ precisely cancels the $\rho_1\rho_1$ term in $\rho_2=\rho_1\rho_1+\Delta\rho_2$, when we take into account the relation $N=1/\lambda_1$.  Altogether our result indeed satisfies the boundary condition (\ref{bc}) at $\Delta t=0$.

Now, for sufficiently large $\Delta t$ and $\chi_t$, we can make a reliable saddle point estimate:
\be\label{saddlew}
w_s = \frac{2\pi^2 i |\chi_t|}{\lambda_1\Delta t}, ~~~ \Delta w_s=\frac{\pi}{\lambda_1}\sqrt{\frac{2}{\Delta t}}
\ee
For the first term in parentheses above, the term that reduced to the delta function at $\Delta t=0$, this saddle point is exact (that term is just exactly a Gaussian integral).  
This saddle point contribution, including both terms in parentheses above, gives
\be\label{saddleanswer}
\Delta\rho_{2,saddle}=\frac{\lambda_1}{2\pi N}e^{-\lambda_1^2\Delta t/4\pi^2}\frac{\pi}{\lambda_1}\sqrt{\frac{2}{\Delta t}}\frac{2\pi^2|\chi_t|}{2\pi^2|\chi_t|-\lambda_1\Delta t}~e^{-v(\chi_t)}e^{-\pi^2\chi_t^2/\Delta t}
\ee
%Now, we see from this that once $\Delta t$ is of order one, say, the Gaussian suppression as a function of $\chi_t$ kills $\Delta\rho_2$ compared to $\rho_1\rho_1$:
%\be\label{rhos}
%\rho_1\rho_1\propto e^{-2\lambda_1 |\chi_t|} \gg \Delta \rho_2 \propto e^{-(\lambda_1 |\chi_t|+\pi^2 \chi_t^2/\Delta t)}
%\ee    
% 
%Preliminary comments on the interpretation:  it may be that the crossover in $\Delta t$ between the non-factorized and factorized forms for $\rho_2$ may be determined by when the residue contribution $w=i$ is suppressed by the $e^{-\frac{w^2\lambda_1^2\Delta t}{4\pi^2}}$ in the integrand.  That would give a criterion on the crossover scale $\Delta t$ that is independent of $\chi_t$.  At the same time, there is nontrivial $\chi_t$ dependence in the result to keep track of to see what effect it has.
%This may also help with the CMB version, to understand the structure of the approximately independent patches on the tail at CMB scales.  

In fact, we can derive a closed form expression for our result as follows.
We want to do the integral:
\be
I=\int_{-\infty}^{\infty}dw e^{-aw^2+2iabw}\left(1+\frac{i}{w-i}\right) = \int_{-\infty}^{\infty} dw e^{-b^2a -a(w-ib)^2} \left(1+\frac{i}{w-i}\right)
\ee
In order to turn the $(w-ib)^2$ factor in the exponent into $w^2$ we can draw a rectangle in the complex plane with sides the real axis and the line $z=bi$. Assume for now that $b<1$ so that the pole is outside the rectangle. Then the two integrals along the two lines are equal:
\be I=\int_{-\infty}^{\infty} dw e^{-b^2a -aw^2} \left(1+\frac{i}{w-(1-b)i}\right) \ee
We need the result:
\be \int_{\infty}^{\infty}dw \frac{e^{-aw^2}}{w-bi} =  bi\int_{\infty}^{\infty}dw \frac{e^{-aw^2}}{w^2+b^2} = \textrm{sign}(b)i e^{ab^2}\pi \textrm{erfc}(\sqrt{a}|b|) \ee
Then,
\be I = e^{-b^2a}\left[\sqrt{\frac{\pi}{a}}-\pi e^{a(1-b)^2}\textrm{erfc}(\sqrt{a}(1-b))\right] \ee
For $b>1$, we have to also include the residue of the pole ($=ie^{a-2ab}$), as well as take into account that $1-b$ is negative:
\be e^{-b^2a}\left[\sqrt{\frac{\pi}{a}}+\pi e^{a(1-b)^2}\textrm{erfc}(-\sqrt{a}(1-b))\right] -2\pi e^{a-2ab}\ee
The two expressions are in fact equal because $\textrm{erfc}(-x) = 2-\textrm{erfc}(x)$.

In our case, $a=\frac{\lambda_1^2\Delta t}{4\pi^2}$ and $b=\frac{2\pi^2|\chi_t|}{\Delta t\lambda_1}$, so we find
\be\Delta \rho_2 = e^{-v(\chi_t)}\frac{\lambda_1}{2\pi N}e^{-\frac{\lambda_1^2 \Delta t}{4\pi^2}} e^{-\frac{\pi^2\chi_t^2}{\Delta t}}\left[\sqrt{\frac{4\pi^3}{\lambda_1^2\Delta t}}-\pi e^{\frac{(\lambda_1 \Delta t - 2\pi^2 |\chi_t|)^2}{4\pi^2\Delta t}}\textrm{erfc}\left(\frac{ \lambda_1 \Delta t - 2\pi^2 |\chi_t| }{2\pi \sqrt{\Delta t}}\right) \right] \ee

\end{document}